\documentclass[12pt,a4paper]{amsproc}

\theoremstyle{definition}

\theoremstyle{remark}

\numberwithin{equation}{section}

\usepackage{latexsym}
\usepackage{amsmath,amsfonts,amssymb}
\usepackage{graphicx}
\usepackage{subfigure,color}
\usepackage{float}
\usepackage{pifont}

\begin{document}
\title{Spectra of graphs and semi-conducting polymers}
\author{Philipp Schapotschnikow}
\address{Utrecht University, P.O. Box 80000 \\3508 TA Utrecht, The Netherlands}
\email{P.Schapotschnikow@uu.nl}
\author{Sven Gnutzmann}
\address{School of Mathematical Sciences, University of Nottingham, United Kingdom}
\email{sven.gnutzmann@nottingham.ac.uk}
\date{\today}
\maketitle
\begin{abstract}
  We study the band gap in some semi-conducting polymers with two models:
  H\"uckel molecular orbital theory and the so-called free electron model.
  The two models are directly related to spectral theory on combinatorial
  and metric graphs. Our numerical results reproduce qualitatively experimental results and
results from much more complex density-functional calculations. We
show that several trends can be predicted from simple graph models
as the size of the object tends to infinity.
\end{abstract}

\maketitle

\section{Introduction}
An important pattern in organic chemistry are delocalized electron
systems, also called conjugated $\pi$-electron systems. Most dyes
owe their color to such structures. When the length of a conjugated
electron system becomes large, electric conductivity and
luminescence are possible.\\
This is achieved in practice by chemically connecting a fragment
with delocalized electrons to its copy, and repeating this procedure
many times. The original fragment is called \textit{monomer}; if the
number of copies is definite and rather small (up to few tenths),
the resulting product is called \textit{oligomer}; if the number of
segments is large and indefinite, the resulting product is called
\textit{polymer}. Semi-conducting polymers are the basis for organic
light-emitting diodes (LED's) and transistors \cite{Friend1999};
oligomers can be used in photovoltaic devices \cite{Ouali1999}.

An important open question about these chemical structures is
whether and how it is possible to extrapolate the electronic
structure of oligomers with increasing number $m$ of segments to
that of the polymer. The asymptotics of the band gap $\triangle
E(m)$ of oligomers with length $m$ increasing to infinity is subject
to ongoing debate
\cite{Daehne1971,Gierschner2007,Hutchison2003,Lewis1939,Luo1998,
Onipko1997,Zade2006}. It is known both from experiments and quantum
chemical calculations that the convergence behavior of the band gap
of oligomers with increasing length to the band gap of the polymer
often deviates from $\triangle E (m) - \triangle E (\infty)\propto
1/m$. One might expect the latter relation based on the behavior of
a linear chain of length $m$ in the limit $m \rightarrow \infty$.
The latter can be calculated straightforwardly and has been observed
experimentally \cite{Bredas1983} (the finite chain corresponds to a
C$_{2m}$H$_{2m+2}$ poly-acetylene oligomer,
 PA$m$).

\begin{figure}
  \begin{center}
    \subfigure[\mbox{Benzene}]{
      \includegraphics[height=1cm]{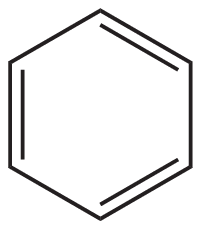}
    \label{fig:benzene}}$\qquad$
    \subfigure[\mbox{Naphthalene}]{
      \includegraphics[height=1cm]{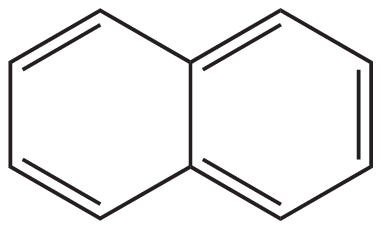}
    \label{fig:naphtalene}} \\
    \subfigure[\mbox{$\beta$-carotene}]{
      \includegraphics[height=1.5cm]{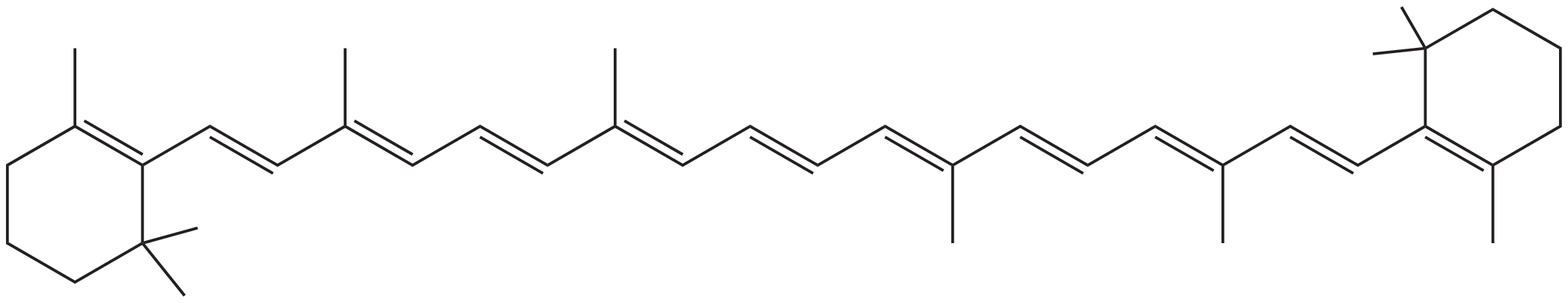}
    }
    \caption{Molecules with conjugated $\pi$-systems. Carbon atoms are conventionally points connected
by bonds. Hydrogen atoms are not displayed. \label{fig:conj_mols}}
  \end{center}
\end{figure}

We will use the following notation for oligomer/polymer molecules:
the polymer is abbreviated by 2-5 letters related to its full name,
e.g.\ PA (poly-acetylene), PPP (poly-para-phenylene). A number $m$
following this abbreviation refers to an oligomer with $m$ copies of
the corresponding monomer, e.g.~PPP$5$ is the para-phenylene
oligomer C$_{30}$H$_{30}$. It consists of 5 connected copies of the
C$_6$H$_6$ monomer benzene.

In this paper we discuss the gap $\triangle E (m)$ and its
convergence in the polymer limit ($m\to\infty$)for quantum chemical
models that reduce to solving an eigenvalue problem on either
combinatorial or metric graphs. Some background on these models will
be given in Section \ref{sec:vb}. In Section \ref{sec:maths} we will
define the spectral problems on graphs and describe how they are
related to oligomers. We will also explicitly solve the simplest
polymer (poly-acetylene in either ring or chain formation) as an
example in Section \ref{sec:ring} and summarize the (well-known)
Floquet-Bloch theory for the infinite polymer in Section
\ref{sec:polylimit}. Eventually we will present and discuss
numerical results for a number of relevant polymers in Section
\ref{sec:example} and compare those to experimental data.

\begin{figure}
  \begin{center}
    \subfigure[PA]{
    \includegraphics[height=0.5cm]{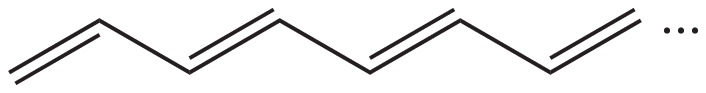}
    }
    \subfigure[PPP]{
    \includegraphics[height=1cm]{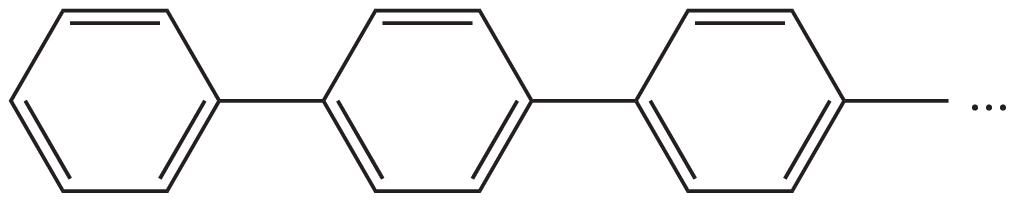}
    }
        \subfigure[PPf]{
    \includegraphics[height=1.5cm]{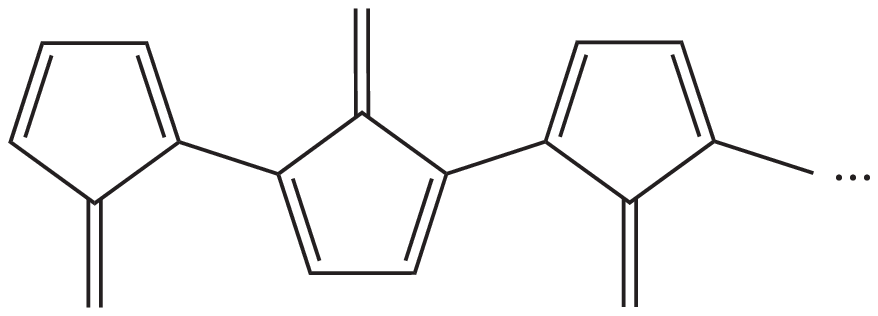}
    }
\caption{Conjugated polymers poly-acetylene, poly-para-phenylene,
poly-pentafulvene. \label{fig:PA_PPP_PPf}}
   \end{center}
\end{figure}

\section{Conjugated double bonds: beyond valence bond theory}
\label{sec:vb}

Conventionally, organic molecules are described by covalent bonds.
According to the \textit{valence-bond (VB) theory} the outer shell
(valence) electrons spatially rearrange themselves such that a pair
of electrons is shared by two atoms while inner shell electrons
remain assigned to individual atoms of the molecule. A single
covalent bond consists of one pair of electrons. Double and triple
bonds between two atoms consist correspondingly of two or three
pairs of electrons shared between the atoms. Recall that one single
carbon atom has four valence electrons and it forms hence four
bonds. Some of these may constitute a double or triple bond. A
sequence of alternating single- and double C--C bonds is called
\textit{conjugated}, see Fig.~\ref{fig:conj_mols} for examples.

The VB assumption fails to explain
the properties of
compounds with conjugated sequences already
on a qualitative level.
 Absorption spectra, luminescence and
conductivity indicate that the spacing
between energy levels in such
molecules is much lower than in similar
non-conjugated counterparts.
This strongly indicates that some electrons
in these molecules are
delocalized over a much larger region
than one single or double
bond, in contrast to the VB theory.

The more general \textit{Molecular Orbital (MO) Theory} overcomes
the shortcomings of the VB theory \cite{Streitwieser1961}. It
postulates that valence electrons recombine and form
\textit{molecular orbitals} to minimize the total energy of the
entire molecule. The following two models have been put forward to
describe conjugated systems in the framework of MO Theory. Both
neglect the repulsion and correlation of electrons. Erich H\"uckel
suggested the following scheme for molecules with alternating double
bonds \cite{Huckel1930,Huckel1931,Huckel1932,Huckel1933}:
\begin{itemize}
\item[i.]
  \textit{Single} covalent
  bonds between adjacent atoms are formed and they determine the
  geometry of the molecule.
\item[ii.]
  The remaining electrons (also called
  {\it $\pi$-electrons}, in a
  conjugated system there is
  one $\pi$-electron per carbon atom)
  and their wave functions
  are subject to energy minimization.
  This reduces
  to an eigenvalue problem on a
  combinatorial graph after
  some further
  simplifying assumptions.
  This spectral problem is referred to as the
  \textit{H\"{u}ckel Molecular
    Orbital (HMO) Theory}.
\end{itemize}
The alternative to the HMO approach is based on the assumption that
the electrons of a conjugated system are strongly bound to the
positively charged carbon-backbone \cite{Schmidt1933}, which is
modeled as a graph \cite{Kuhn1948,Ruedenberg1953}. Interestingly, in
the original formulation Hans Kuhn already considered atoms as
scattering centers for the electron wave functions. This is the
\textit{Free Electron Model (FE)}. This model can be reduced (after
separation of variables) to an eigenvalue problem on a metric graph
for the Schr\"odinger operator without a potential (hence the `free
electron' in the name). \footnote{Arguments of \cite{Ruedenberg1953}
are incorrect for MO's localized at an atom (eigenfunctions of the
3D-Schr\"odinger operator localized at a vertex)
\cite{Kuchment2002}. Such states would correspond according to the
MO Theory to unbound electron pairs and are, therefore, irrelevant
for realistic hydrocarbons.}
 The assumptions of this model are
only plausible for large systems.

For finite
molecules both, the MO and the FE model,
yield discrete energy
levels $E_n$ with corresponding
eigenfunctions (molecular orbitals).
Owing to spin, each of these levels can
be occupied by two electrons
(applying Hund's rule and
the Pauli exclusion principle).
In a system with $M$
alternating double bonds there are
$2M$ $\pi$-electrons. In the
ground state of the system, the
$M$ lowest molecular orbitals are
occupied.
The energy
of the highest occupied
molecular
orbital (\textit{HOMO}) is thus
$E_{HOMO}=E_M$.
The
energy of the lowest unoccupied molecular
orbital (\textit{LUMO}) is $E_{LUMO} =
E_{M+1}$.\\
The difference $\triangle E=E_{LUMO}-E_{HOMO}$ is the smallest
energy package a molecule can absorb. It is directly related to the
color of a material (i.e.~to the highest absorbed wave length) and
to the breakdown voltage of semi-conducting materials (since this is
proportional to the band-gap). In Ref.~\cite{Wennerstrom1985} band
gaps of several semi-conductive polymers have been studied
qualitatively using HMO theory.

The degeneracy of $E_M$ can determine chemical properties of a
compound. If $E_M$ is a multiple, completely occupied energy level,
then the compound is chemically stable and inert. The classical
example is nitrogen. Conversely, if $E_M$ is multiple and not
completely occupied (\textit{\it SOMO}, where ``S'' stands for
``singly''), then the compound is chemically reactive. The classical
example is oxygen. The existence of such orbitals in conjugated
hydrocarbons is related to the question whether 0 is an eigenvalue
of the adjacency matrix of the underlying graph, see \cite[Chapter
8]{Cvetkovic1995} which is dedicated to this topic.

One advantage of the
HMO theory is that absolute
energy levels can be
calculated.
As a consequence it is possible to
investigate the stability
of compounds. Moreover, the
underlying MO theory is not
restricted to
conjugated systems.\\
On the other side, the FE model has the advantage that no parameter
fitting is involved. Moreover, wave functions and electron densities
can easier be calculated in the FE model. The latter information can
then be used to infer statements
 about bond lengths and
effect of heteroatoms \cite{Kuhn1995}.

\section{Mathematical description of polymers by combinatorial and metric graph models}
\label{sec:maths}

For a given molecule with a conjugated fragment and $N_C$ carbon
atoms, it is straightforward to define the underlying graph $G$. The
vertices $v_i$ (where $i=1,2,\dots,N_C$) of the graph are the carbon
atoms and the undirected bonds $b$ of the graph are the chemical
bonds between them. For the undirected bond connecting the vertices
$v_i$ and $v_j$ we write $b_{ij}\equiv b_{ji}$. Note that on the
graph $G$ \textit{all bonds are simple}, disregarding  their
chemical multiplicity. We write $i \sim j$ if the vertices (atoms)
$v_i$ and $v_j$ are connected. The undirected (metric) graph is
fully characterized by its $N_C \times N_C$ symmetrical bond matrix
$B$, where $B_{i,j}=B_{j,i}$ is the bond length between the two
carbon atoms (vertices) $v_i$ and $v_j$ if they are connected, and
$0$ otherwise. In the following all bonds in a purely hydrocarbon
system are assumed identical, unless the opposite is stated. By
replacing all bond lengths in $B$ with unity one obtains the $N_C
\times N_C$ connectivity matrix $C$.

HMO theory in its
original formulation solves the
eigenvalue problem
\begin{equation}
  Cu= \lambda u.
\end{equation}
Since $C$ is a real symmetric matrix, it
has $N_C$ real eigenvalues
$\lambda_1,\dots,\lambda_{N_C}$. The
electronic energy levels $E_n$
in a molecule are related to the
eigenvalues $\lambda_n$ of the
connectivity matrix by
\begin{equation}
  E_n = \lambda_n \beta - \alpha, \quad n=1,\dots,N_C
\end{equation}
where $\beta$ and $\alpha$ are constant parameters with units of
energy. The value of $\alpha$ is not relevant for pure hydrocarbon
compounds since it is just a constant shift. The parameter $\beta$
is defined as a resonance integral, but is usually fitted to
experimental data. In order to keep results consistent with the FE
model, we choose $\beta=3.05$eV as in Ref.~\cite{Taubmann1992}. It
is important to note that the parameters $\alpha$ and $\beta$ are
not potentials in the physical sense, and that the eigenvectors
$u_n$ are not the values of the wave function at the vertices.

To formulate the FE model, we need to define a real coordinate $0\le
x_{ij}\le B_{ij}$ along each undirected bond $b_{ij}$. For $i<j$ we
set $x_{ij}=0$ at one end of the bond where it is connected to the
vertex $v_i$ and $x_{ij}=B_{ij}$ at the other end connected to
$v_j$. We will also use the notation $x_{ij}= v_i$ for the value of
$x_{ij}$ at vertex $v_i$. The metric graph is quantized by defining
a self-adjoint Schr\"odinger operator for a wave function $\Phi(x)$
on the metric graph.\\
This implies
 that
the wave function
$\Phi(x)$ must satisfy the
one-electron free Schr\"odinger
equation on each
bond $b_{ij}$
\begin{equation}
  -\frac{\hbar^2}{2 m_e} \frac{d}{dx^2_{ij}} \Phi (x_{ij}) = E \Phi (x_{ij}) \label{eq:Schro_eq}
\end{equation}
with Plank's constant $\hbar$,
electron mass $m_e$ and
energy $E$.\\
To obtain a self-adjoint Schr\"odinger operator one has to add
certain conditions on the wave functions at the vertices
\cite{gnutzmann2006,kostrykin1999,kottos1999}. We will impose the
Neumann conditions (also known as Kirchhoff conditions) that the
wave function is continuous at each vertex
\begin{align}
  \text{for all $j \sim i$}\;\lim_{x_{ij}\rightarrow v_i}
  \Phi(x_{ij})=& \Phi(v_i)  \label{eq:cont_cond} \\
\intertext{and, additionally that the
sum over all outgoing derivatives of the wavefunctions on all bonds connected
to a vertex vanishes}
\left.
  \sum_{j: j  \sim i} w_j^i \frac{d}{dx_{ij}}\Phi(x_{ij})\right|_{x_{ij}=v_i} =& 0   \label{eq:Kirchhoff}
\end{align}
where $w_j^i=1$ if $i<j$ and
$w_j^i=-1$ otherwise.

Solving the Schr\"odinger Equation~(\ref{eq:Schro_eq}) with vertex
conditions (\ref{eq:cont_cond}-\ref{eq:Kirchhoff}) we obtain
discrete energy levels $E_n$ and wave functions (MOs) $\Phi_n$.

Additional potentials on the bonds have been introduced in
Eq.~(\ref{eq:Schro_eq}) to improve the agreement with experimental
values \cite{Kuhn1948,Ruedenberg1953}. One may also add vertex
potentials\footnote{A vertex potential is equivalent to changing the
inner condition \eqref{eq:Kirchhoff} at a vertex while keeping the
continuity condition \eqref{eq:cont_cond}} to get a more realistic
description (see \cite{gnutzmann2006,kottos1999}). Any such
generalization adds some fitting parameters. In this work we are
mainly interested in more fundamental questions concerning the
convergence behavior for general polymers and we thus stick to the
simplest model and no fitting parameters.

We scale the
bond lengths of the graph to the
average C--C bond length of
$L_{C-C}= 1.4$\AA\ and $\frac{\hbar^2}{2 m_e}$ to $1$. The
eigenvalues $\mu_n$ of the scaled problem
correspond to electronic
energy levels as $E_n = \mu_n \epsilon$ with
$\epsilon=\frac{\hbar^2}{2m_e L_{C-C}^2}=1.95$eV.

In order to find the eigenvalues numerically, one has to solve the
secular equation (see \cite{gnutzmann2006,kottos1999} for details).
However, throughout this work we assume equal bond lengths. Then the
following identity is valid \cite{Pankrashkin2006}. Define the
diagonal valency matrix $V$ (i.e. $V_{i,i}$ is the number of bonds
at the vertex $v_i$). If $\mu$ is an eigenvalue of the (scaled) FE
problem, then $\tilde\mu = \cos \sqrt\mu $ is an eigenvalue with the
same multiplicity of the generalized eigenvalue problem
\begin{equation}
  Cu = \tilde\mu Vu .
\end{equation}
%with the connectivity matrix $C$ and diagonal.

\begin{figure}
  \begin{center}
    \includegraphics[height=2.5cm]{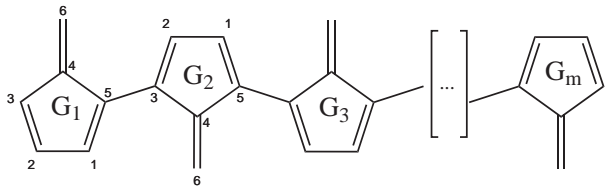}
    \caption{Construction of the oligomer PPf$m$, chemical notation is used to represent the graph.
    The fifth carbon atom of the first monomer in the chain is connected to the third atom of the second one etc. In the graph notation this means
    $v_5^1 \sim v_3^2$; the index $b$ is $5$ and the index $e$ is $3$.  \label{fig:constr_oli}}
  \end{center}
\end{figure}

We adopt the following notation for oligomers and polymers, see also
Fig.~\ref{fig:constr_oli}. A monomer $X\equiv X1$ is described by a
graph that we will call $G_1$. The number of carbon atoms in the
monomer will be denoted by $N_C$ and is equal to the number of
vertices in $G_1$, and $N$ will be used for the number of double
bonds in the monomer. Usually, $N_C=2N$. The monomer has two atoms
at which bonds to other monomers of the same type are created. Let
these atoms be represented by the vertices $v_b,v_e$, so that two
consecutive monomers are connected by a bond between the atom $b$ of
the first and $e$ of the second monomer. Let us now consider $m$
copies $G_1, G_2, G_3,\dots, G_m$ of the monomer $G_1$. Denote the
vertex $v_i$ on $G_a$ by $v_i^a$. We can now construct the graph
$G^m$ that represents the oligomer $Xm$ as a union $G^m =
\bigcup_{a=1}^m G_a$ with extra bonds between $v^a_b$ and
$v^{a+1}_e,\ a=1,\dots,m-1$. Obviously one has $G_1=G^1$. The band
gap of $G^m$ is
\begin{equation}
\triangle E(m) = E_{M+1} - E_{M}
\end{equation}
where $M=Nm$ is the number of double bonds
on the oligomer.

The above construction assumes that adjacent monomers are connected
via a single bond between two atoms. We only consider polymers of
this type here. Note that the construction can easily be generalized
to incorporate many bonds between adjacent monomers.

\subsection{Examples: rings and chains}
\label{sec:ring} Consider a ring C$_{2m}$H$_{2m}$ consisting of
$M=m$ alternating double bonds. It has $N_C=2$  carbon atoms per
"monomer", each with $N=1$ (trivially) alternating double bond. The
HMO-eigenvalues are
\begin{equation}
\lambda_n =  2 \cos (\frac{n\pi}{m}), \quad n=1,2,\dots,2N
\end{equation}
while the FE eigenvalues are
\begin{equation}
\mu_n = \frac{n^2 \pi^2}{m^2},\quad n=0,\pm 1,\pm 2,\dots
\end{equation}
We see that both models predict qualitatively the same energy
scheme. The ground state is simple, higher states are double (except
for $\lambda_{2m}$, which is not important). Thus, $E_m$ and
$E_{m+1}$ are both double. For odd $m$, $E_{HOMO}$ is completely
filled, yielding an increased chemical stability, while for even $m$
it is not complete leading to an increased reactivity. This is the
famous H\"{u}ckel rule.

Consider now a chain C$_{2m}$H$_{2m+2}$ with $M=m$ alternating
double bonds {\it(polyenes)}. The limit $m\to \infty$ is the most
trivial example for a conjugated polymer and is called
poly-acetylene (PA). It was the first polymer for which electric
conductivity was shown and studied.

The HMO and FE eigenvalues are
\begin{align}
\lambda_n &=  2 \cos \left( \frac{n\pi}{2m+1}\right), \quad n=1,2,\dots,2m \\
\mu_n     &=  \frac{n^2\pi^2}{(2m-1)^2}, \quad n=0,1,2,\dots
\end{align}
respectively \footnote{ In the case of the free electron model
(metric graphs) it seems rather unnatural to assume that the MOs in
a linear chain are limited by the end atoms.  We thus add dangling
bonds at $v_e^1$ and $v_b^m$, so that the total length of the MO
becomes $2N+1$ \cite{Ruedenberg1953,Taubmann1992},
  yielding $\lambda_n=\frac{n^2\pi^2}{(2m+1)^2}$
  and
  $\triangle E_{FE}(m) =
  \frac{\epsilon \pi^2}{2m+1}$.}.
Now all energy levels a single. The HOMO-LUMO
difference in HMO is
\begin{equation}
\triangle E_{HMO}(m) = E_{m+1}-E_{m}= -4\beta
\sin\left(\frac{\pi}{4m+2}\right)
\end{equation}
and in FE
\begin{equation}
\triangle E_{FE}(m) = E_{m+1}-E_{m}= \epsilon \pi^2
\left(\frac{2m+1}{(2m-1)^2}\right).
\end{equation}

We see that both models predict $\triangle E (\infty) = 0$ and
$\triangle E (m)-\triangle E (\infty) \propto \frac{1}{m}$ for large
$m$. In both models, the discrete electronic levels become one
single band in the $m\to\infty$ limit. In practice, the band gap of
PA is significant ($1.8$eV), but the convergence behavior $\triangle
E (m)-\triangle E (\infty)$ is almost linear in $1/m$ with the slope
close to the FE and HMO case, see Section~\ref{sec:example}.

\subsection{Polymer limit} \label{sec:polylimit}

In the limit $m\to\infty$  the underlying graph $G^\infty$ is periodic
 and Floquet-Bloch Theory can be applied
\cite{Avron1994,Harrison2007,Kuchment1991,Kuchment2005,Kuchment2007,Kuchment2006,Rabinovich2007}.
Consider the HMO eigenvalue problem (the FE case is analogous and
will not be discussed in detail)
\begin{equation}
C_{G^\infty} \Psi = \lambda \Psi.
\end{equation}
 All physically relevant eigenvector candidates (bound
states) can be found as a product of a periodic function and a phase
shift:
\begin{equation}
  \Psi(v_i^{j+1})=e^{ik}\Psi(v_i^j),
  \label{eq:FlBl}
\end{equation}
where we introduced the quasi-momentum $k\in \left[-\pi,
\pi\right]$. For a multiple eigenvalue $\lambda$, Floquet-Bloch
eigenfunctions span the eigenspace. The eigenvalue problem on
$G^\infty$ reduces to the problem on a single monomer
\begin{equation}
  (C+F(k))\Psi = \lambda(k) \Psi
\end{equation}
where $F(k)$ is a $N_C\times N_C$ matrix with only two non-zero
entries at the connection between monomers  $F_{eb}(k)=e^{ik}$ and
$F_{be}(k)=e^{-ik}$. Since $C+F(k)$ is a Hermitian matrix, it has
$N_C$ real eigenvalues $\lambda_1(k),\dots,\lambda_{N_C}(k)$. Each
of the branches $\lambda_r(k)$ is a continuous function of $k$. The
set $\bigcup_{0\le k \le \pi}\lambda_r(k)$ is the $r$th band. The
$N$th band is called the {\it valence band} while the $(N+1)$th band
is the {\it conduction band}.

The spectrum of $C_{G^\infty}$ is the union of all bands
\begin{equation}
  \sigma(C_{G^\infty})=\bigcup_{0\le k \le \pi} \bigcup_{r=1,\dots,2N}
  \lambda_r(k).
\end{equation}
Due to the one-dimensional periodicity and the fact that the
monomers are connected by one single bond, the extrema of
$\lambda_r(k)$ are always at $k=0$ or $k=\pi$. Consider the
characteristic polynomial $\xi$ of $C+F(k)$ as a function of $k$ and
$\lambda$
\begin{equation}
\xi(k,\lambda) = \det(C+F(k)-\lambda I_{N_C})
\end{equation}
where $I_{N_C}$ is the unit matrix. Eigenvalues are given by
$\xi(k,\lambda)=0$, and hence the Implicit Function Theorem can be
applied: if $\frac{\partial \xi}{\partial \lambda} \ne 0$ in a
neighborhood of $\tilde k,\tilde\lambda$ with $\xi(\tilde
k,\tilde\lambda)=0$, then $\lambda(k)$ is a well-defined,
differentiable function in this neighborhood with
\begin{equation}
\frac{d \lambda}{d k} = - \left(\frac{\partial \xi}{\partial
\lambda} \right) \times \left( \frac{\partial \xi}{\partial
k}\right).
\end{equation}
One easily finds that
\begin{multline}
\frac{\partial \xi}{\partial k} = i e^{ik} (-1)^{e+b} M_{e,b} - i
e^{-ik} (-1)^{e+b} M_{b,e} =\\
 -2\, (-1)^{e+b}{\rm Im}(e^{ik} M_{e,b})
 = {\rm Im}\left[ e^{ik} \left( m_0(\lambda)+m_1(\lambda)e^{-ik} \right)
 \right] = m_0(\lambda) \sin k
 \label{eq:dxidk}
\end{multline}
where $M_{e,b}$ is the $e,b$-minor of $C+F(k)-\lambda I_{N_C}$, i.e.
the determinant of the matrix after deleting the $e$th row and $b$th
column; $m_0,m_1$  are some polynomials in $\lambda$. The first
equation follows from the rule for determinant derivatives; the
second one from the fact that $M_{e,b} = \overline{M}_{b,e}$; the
third one because $M_{e,b}$ has only one entry which depends on $k$;
and the last one because $m_1(\lambda)$ is real. We obtain therefore
\begin{align}
\xi(k,\lambda) &= p_0(\lambda)\cos(k) + p_1 (\lambda),
&\frac{\partial \xi}{\partial \lambda} &= p'_0(\lambda) \cos(k) +
p'_1(\lambda) \label{eq:xi}
\end{align}
with some polynomials $p_0$ ($=-m_0$) and $p_1$. Obviously,
$\lambda(k)=\lambda(-k)$ and $\lambda$ determines $k\in [0,\pi]$
uniquely iff $p_0$ does not vanish.
 Now three cases are possible for an eigenvalue $\tilde\lambda$:
\begin{enumerate}
\item $p_0(\tilde\lambda)=p_1(\tilde\lambda)=0$. Then $\xi(k,\tilde\lambda)$
does not depend on $k$, and $\tilde\lambda$ is a degenerate band
$\lambda(k)=\tilde\lambda$. And, conversely, if $\tilde\lambda$
belongs to a non-degenerate band, then $p_0(\tilde\lambda)\ne 0$.
\item $\frac{\partial \xi}{\partial \lambda}(\tilde\lambda) = 0$.
This means that $\tilde\lambda$ is a multiple eigenvalue, i.e.
belongs to two (or more) bands. In this case, the two bands join.
Assume namely, that two non-degenerate bands overlap, i.e. there
exists an interval $(\lambda_-,\lambda_+)$ of ``double''
eigenvalues. Then, according to Eq.\eqref{eq:xi}, almost everywhere
on this interval would hold $\cos(k)=p_1(\lambda)/p_0(\lambda)$ and
$p'_0(\lambda)p_1(\lambda)+p'_1(\lambda)p_0(\lambda)=0$. Therefore,
both $p_0$ and $p_1$ must be constant, which is only possible for
empty graphs.
\item Otherwise, $p_0(\tilde\lambda) \ne 0$ on the entire band, and the
extrema of $\lambda(k)$ within this band are given by the zeros of
Eq.\eqref{eq:dxidk}, i.e. integer multiples of $\pi$.
\end{enumerate}
 Thus, in all cases the band edges
are contained in the eigenvalues of $C+F(0)$ and $C+F(\pi)$
($e^{ik}= \pm 1$). Note that in the second case no statement is made
for degenerate bands. Remarkably, if $F(k)$ has more than two
non-zero entries, i.e.\ the monomers are connected by more than one
bond, then the arguments of Eq.\eqref{eq:dxidk} become incorrect,
and a counterexample to the statement about band edges can be found
in \cite{Harrison2007}.

To study and compare spectral bands of oligomers and polymers, it is
useful to define for a real number $x$ the spectral counting
function as the number of eigenvalues per monomer, which are smaller
than $x$. For an oligomer this is given by
\begin{equation}
\sigma^m(x) = \frac{\{n | \lambda_{n-1}< x \le \lambda_{n}\}}{m}.
\end{equation}
where the eigenvalues $\lambda_0,\lambda_1,\dots$ are put in
increasing order and $\lambda_{-1}\equiv -\infty$. The spectral
counting function of the polymer is obtained
by taking the limit $m\to \infty$.\\
Intervals where $\sigma$ increases continuously correspond to bands;
intervals where it is constant correspond to band gaps. Jumps are
degenerated bands: e.g.~if $G_1$ has an eigenvalue $\Lambda$ with an
eigenstate which vanishes at the vertices $v_b$ and $v_e$ then the
oligomer will have an $m$-fold degenerate eigenvalue with
quasi-momentum $k=0$ and
$\lambda_1(0)=\lambda_2(0)=\dots=\lambda_m(0)=\Lambda$ with
eigenstates which are localized on one monomer. Such eigenstates do
not satisfy Eq. \eqref{eq:FlBl}, but lie in the space spanned by
Floquet-Bloch eigenfunctions.
% e.g.~ eigenvalues with multiplicity $m$ with
%localized
%eigenfunctions.

\begin{figure}
\begin{center}
\includegraphics[width=10cm]{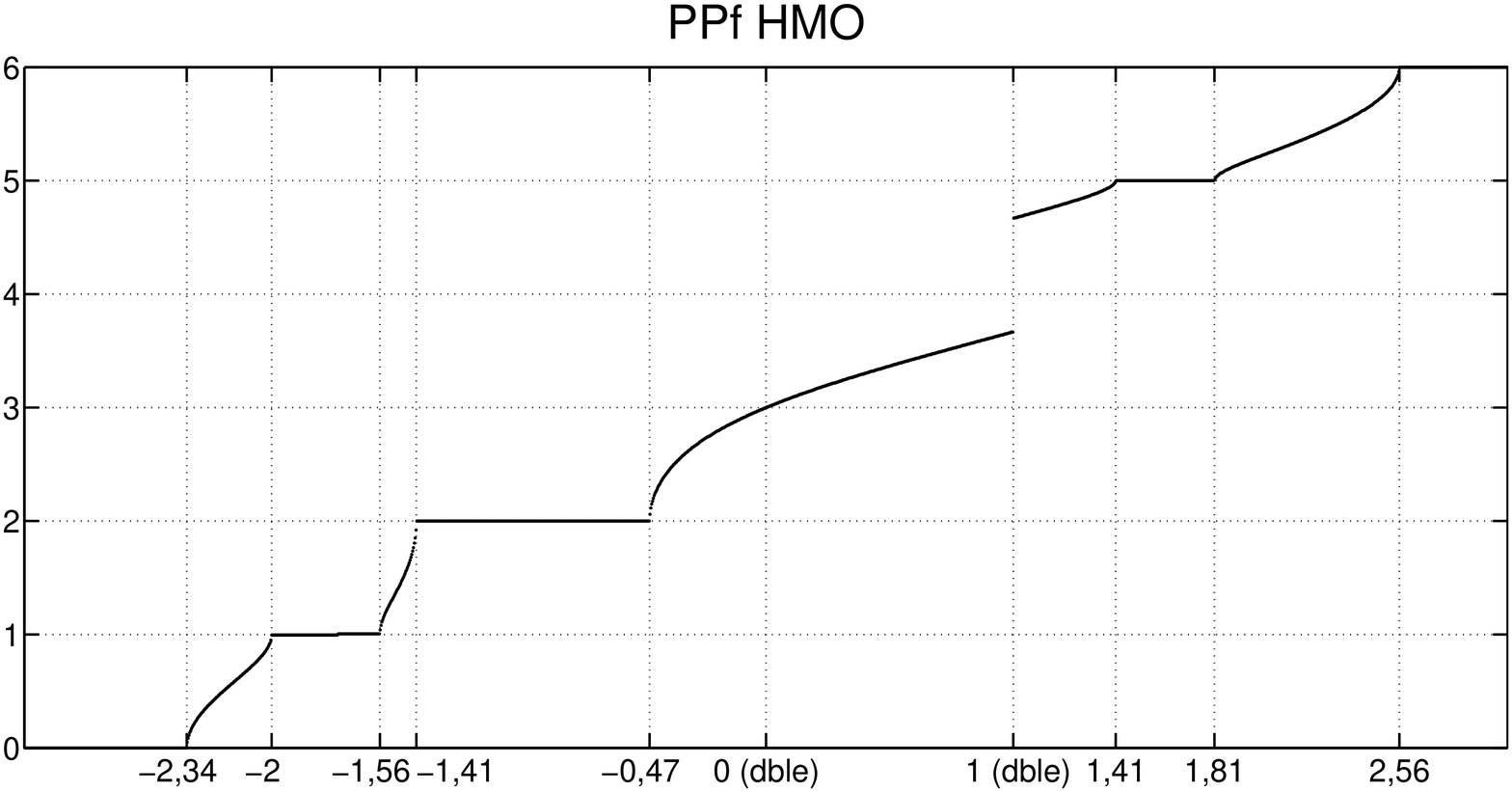}
\caption{$\sigma^{200}_{\rm HMO}$ and Floquet-Bloch Theory of PPf. The
twelve band edges arranged in increasing order $-2.34$, $-2$,
$-1.56$, $-1.41$, $-0.47$, $0$, $0$, $1$, $1$, $1.41$, $1.81$,
$2.56$ are displayed on the horizontal axis. \label{fig:PPf}}
\end{center}
\end{figure}

Consider PPf, see Fig.~\ref{fig:constr_oli}. The matrices $C$ and
$F(k)$ are
\begin{align}
C&=\left(
      \begin{array}{cccccc}
        0 & 1 & 0 & 0 & 1 & 0 \\
        1 & 0 & 1 & 0 & 0 & 0 \\
        0 & 1 & 0 & 1 & 0 & 0 \\
        0 & 0 & 1 & 0 & 1 & 1  \\
        1 & 0 & 0 & 1 & 0 & 0 \\
        0 & 0 & 0 & 1 & 0 & 0 \\
      \end{array}
    \right)
;  &F(k)&=\left(
            \begin{array}{cccccc}
              0 & 0 & 0 & 0 & 0 & 0 \\
              0 & 0 & 0 & 0 & 0 & 0 \\
              0 & 0 & 0 & 0 & e^{ik} & 0 \\
              0 & 0 & 0 & 0 & 0 & 0 \\
              0 & 0 & e^{-ik} & 0 & 0 & 0 \\
              0 & 0 & 0 & 0 & 0 & 0 \\
            \end{array}
          \right).
\end{align}
 The calculated $\sigma^{200}_{\rm HMO}$ is shown
in Fig.~\ref{fig:PPf} together with the band edges from
Floquet-Bloch Theory. The first two bands are $[-2.34,-2]$ and
$[-1.56,-1.41]$. Since $0$ and $1$ appear twice, there are no gaps
between the third, fourth and fifth bands. The last band is then
$[1.81,2.56]$. Knowing that $0$ is a double eigenvalue of $C+F(0)$
while $1$ is a simple eigenvalue of $C+F(0)$ and $C+F(\pi)$, we
conclude that there is a jump at $1$ and no jump at $0$.

\section{Examples and comparison with experiment}
\label{sec:example}

It is important to note that experimentally determined band gaps
systematically differ from the ones predicted by simple theories.
The differences are due to {\it (i)} intermolecular phenomena such
as electronic repulsion, electron correlation, spatial geometry,
non-conjugated side chains and {\it (ii)} intermolecular effects
such as solvent or solid-state shifts and impurities. These effects
are approximately constant ($\approx 1$eV) for a given sequence of
oligomers at the same experimental conditions \cite{Gierschner2007}.
 The band gap of a (real) material is influenced by the
environment, therefore different measurements on the same compound
may give slightly different values depending on experimental
conditions.

\begin{figure}
\begin{center}
\mbox{\hspace*{-1cm}
  \includegraphics[width=0.6\textwidth]{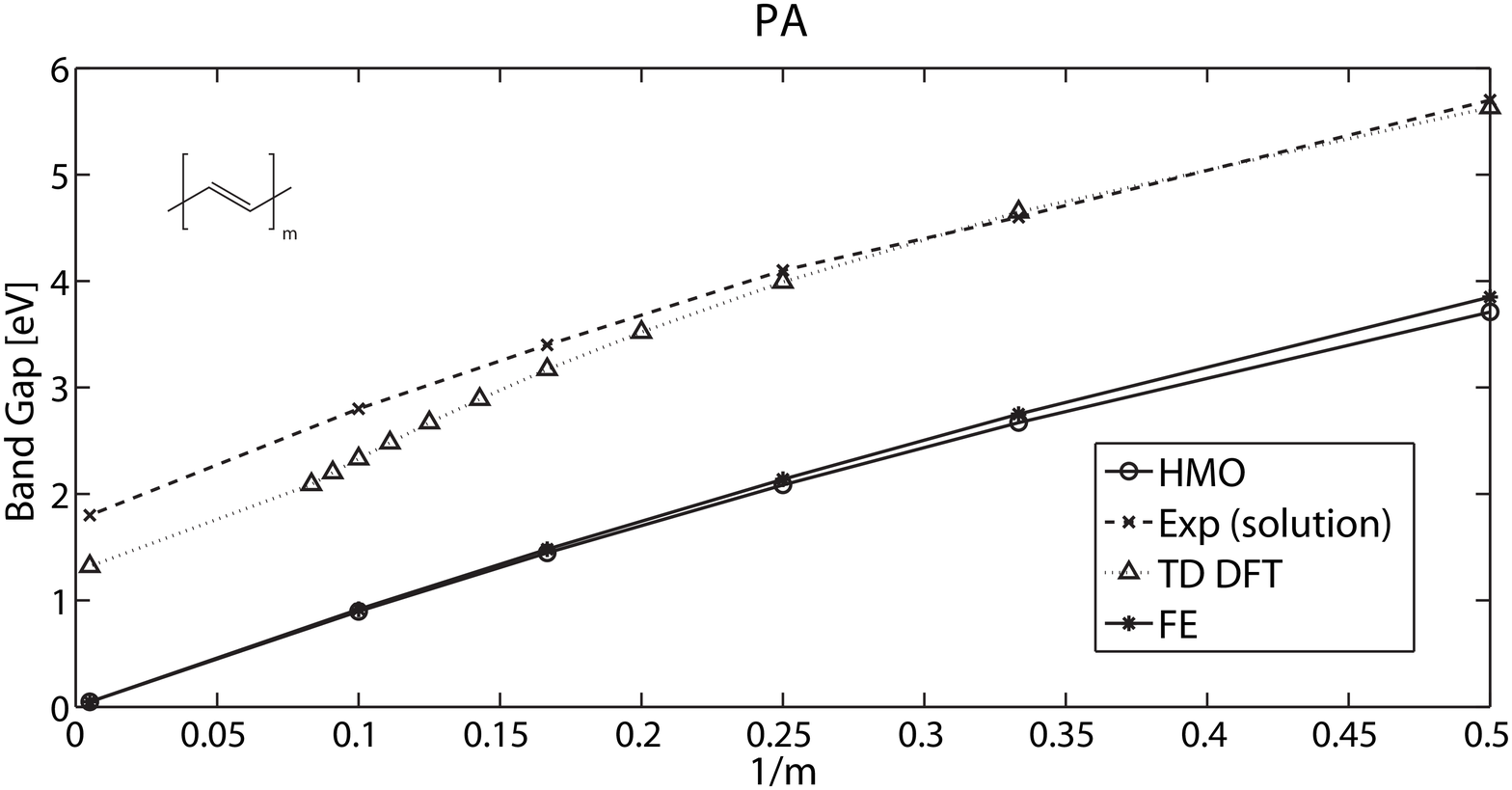}
  \includegraphics[width=0.6\textwidth]{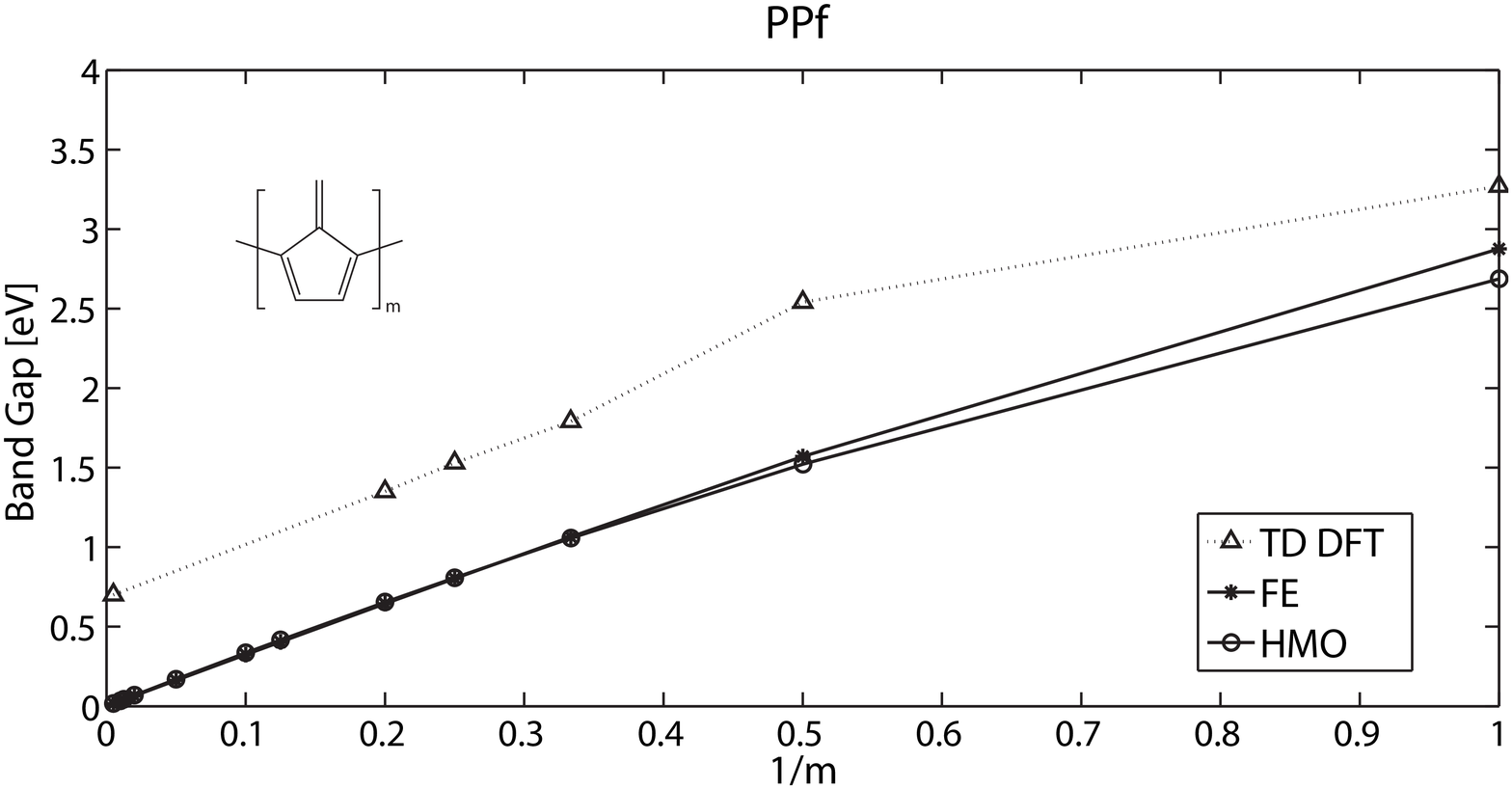}
  \hspace{-1cm}}
\mbox{\hspace*{-1cm}
  \includegraphics[width=0.6\textwidth]{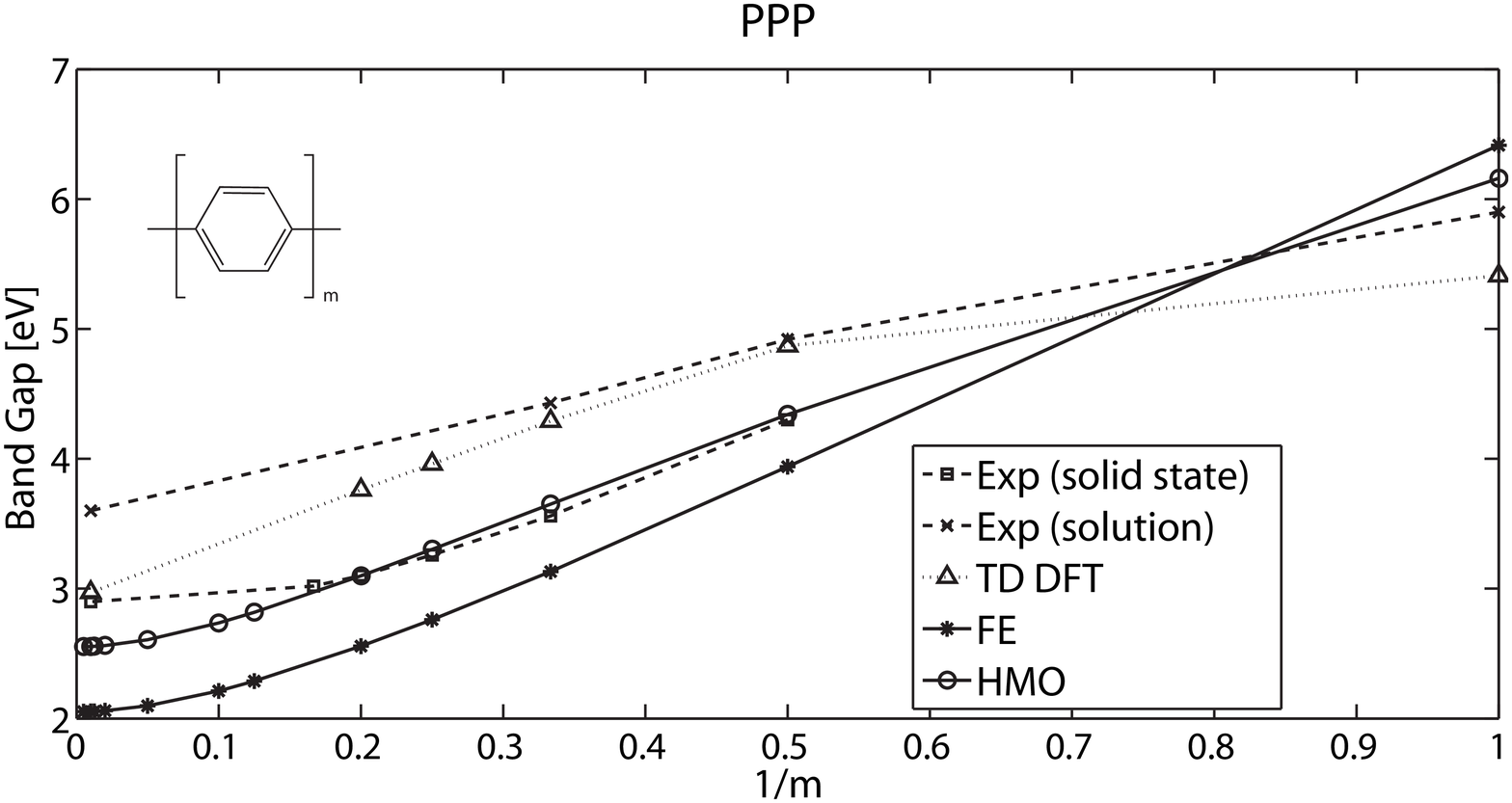}
  \includegraphics[width=0.6\textwidth]{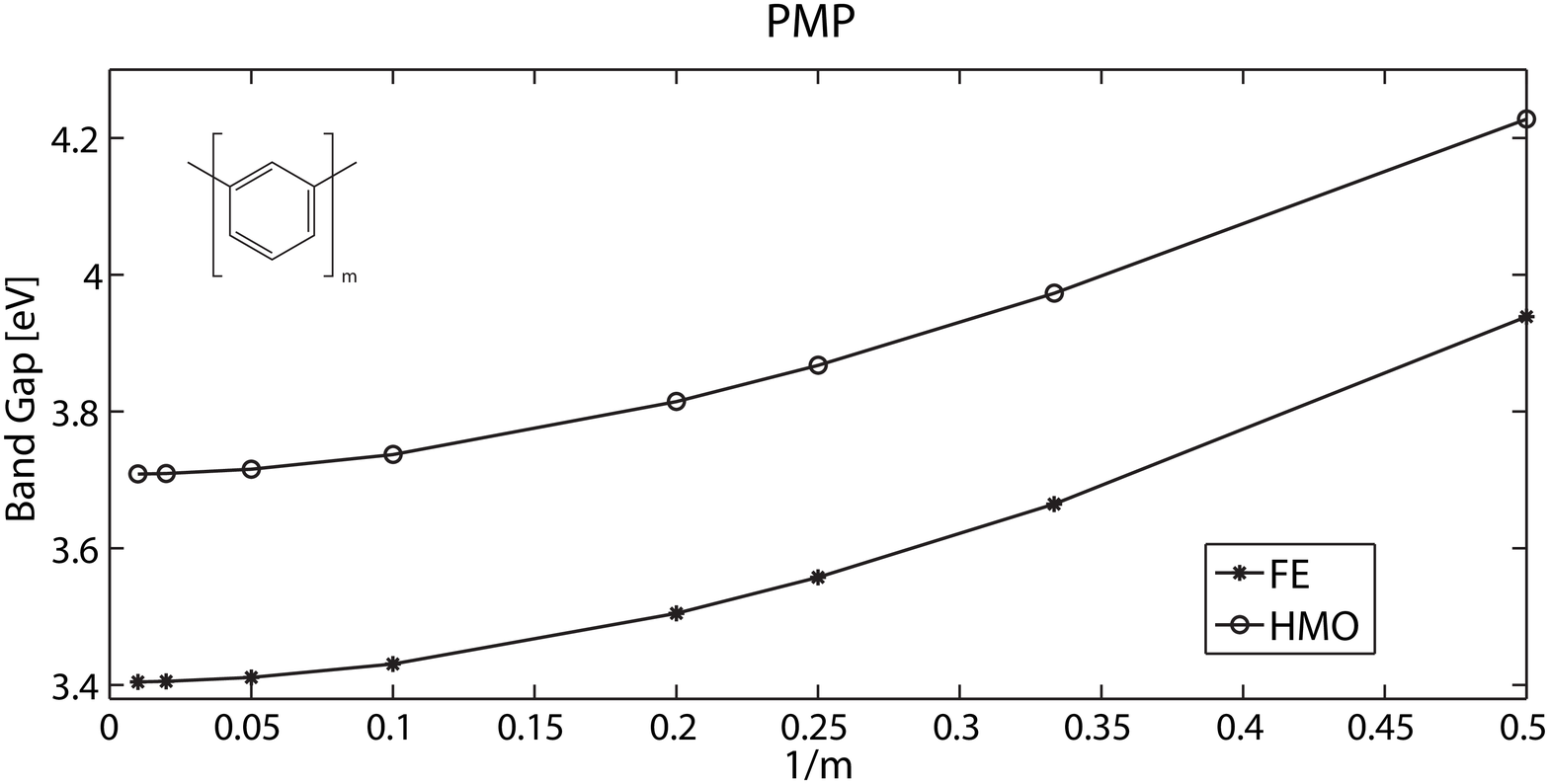}
\hspace{-1cm}} \mbox{\hspace*{-1cm}
  \includegraphics[width=0.6\textwidth]{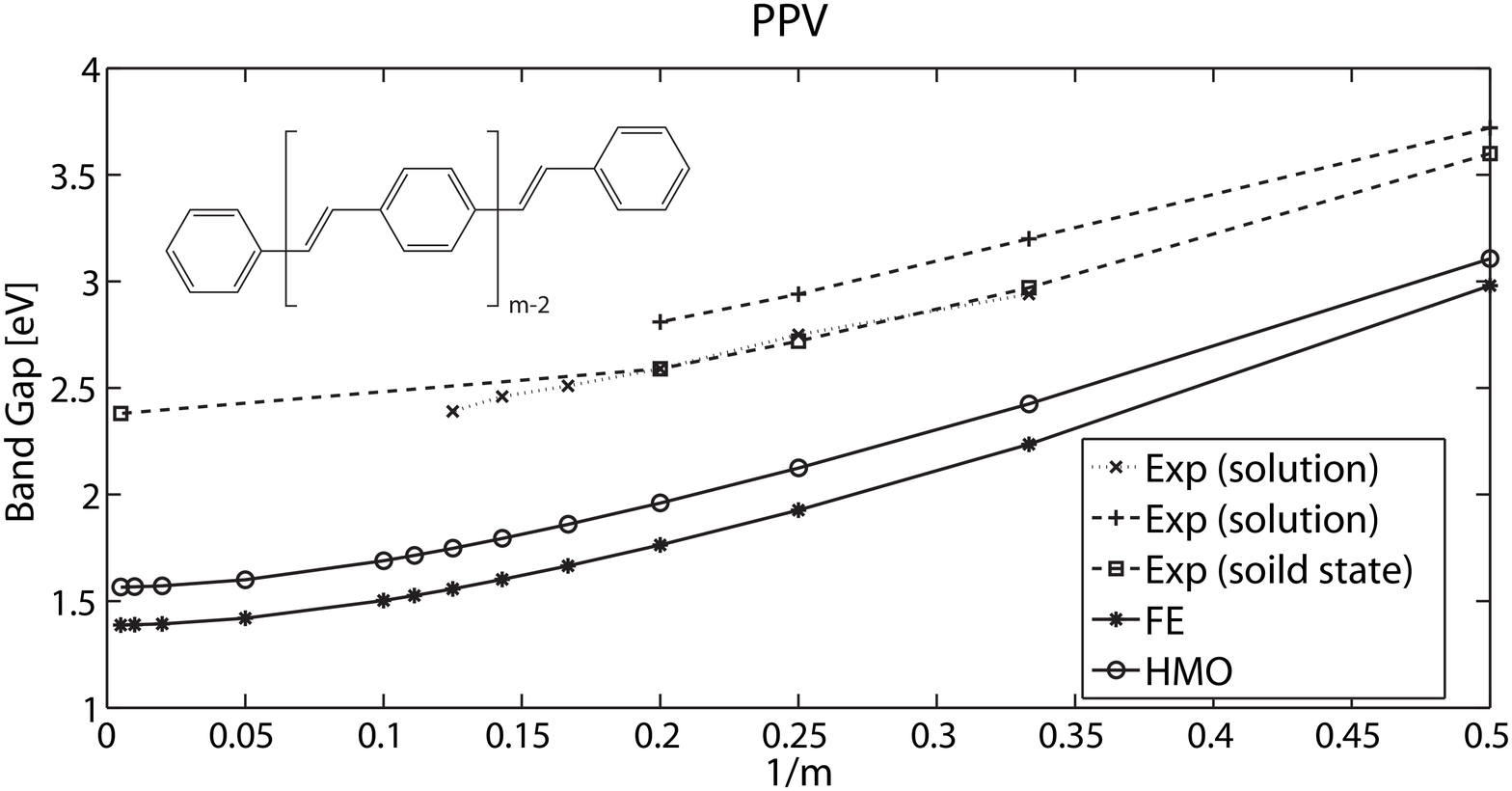}
  \includegraphics[width=0.6\textwidth]{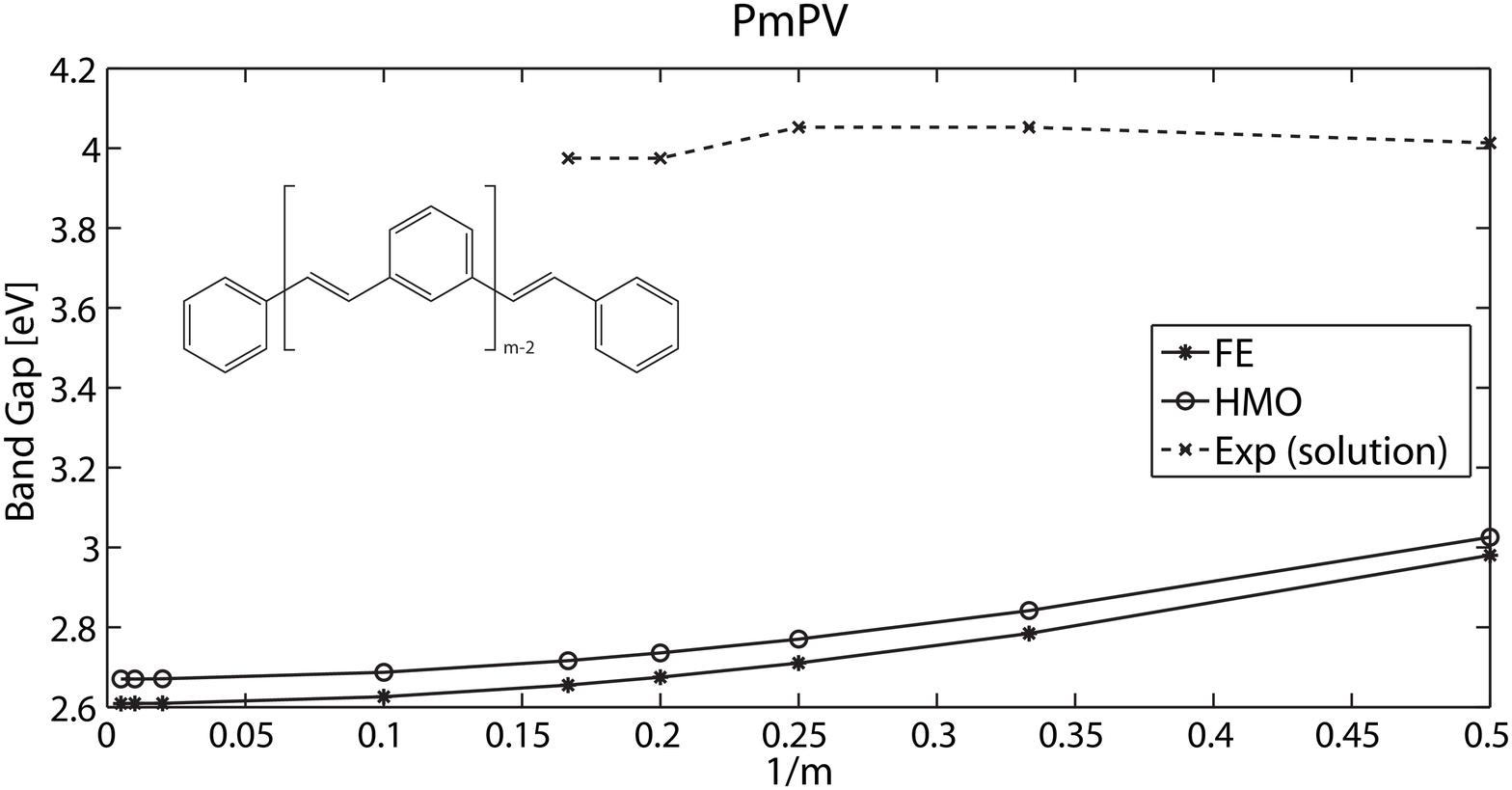}
\hspace{-1cm}}
\end{center}
\caption{Band gaps of oligomers estimated by HMO and FE models
compared with experimental values. On the horizontal axis is the
inverse number of monomers. HMO-values are represented by circles;
FE-values by stars. Sources: all TD DFT calculations \cite{Ma2002};
PA \cite{Bredas1983}; PPP
\cite{Matsuoka1991},\cite{Nijegorodov2000}; PPV
\cite{Gebhardt1999},\cite{Gierschner2002},\cite{Gierschner2007};
PmPV \cite{Gregorius1992}.  \label{fig:BG_limit}}
\end{figure}

We consider the following polymers: PA ($N_C=2$, $N=1$), PPf
(poly-pentafulvene, $N_C=6 $, $N=3$), PPP ($N_C=6$, $N=3$), PMP
(poly-meta-phenylene, $N_C=6$, $N=3$), PPV (poly-phenylene-vinylene,
$N_C=8$, $N=4$) and PmPV (poly-meta-phenylene-vinylene, $N_C=8$,
$N=4$), see also Fig.~\ref{fig:PA_PPP_PPf}. The first one is the
alternating chain and has already been discussed in Section
\ref{sec:ring}. The second one, PPf represents the large number of
$5$-ring based polymers. To the best of our knowledge, it is subject
to theoretical studies \cite{Hong2000,Pranata1988} as a possible
low-band-gap polymer, but has not been synthesized yet. The other
four are based on benzene (6-ring) as monomer but connected
differently. PMP has a $1-3$ connection ({\it meta}) whereas PPP has
a $1-4$ connection ({\it para}). PPV and PmPV have additionally two
carbon atoms between the rings, i.e. the connection length is $3$
and $N=4$. Note that PPP$2$=PMP$2$ and PPV$2$=PmPV$2$.

As we can see from Fig.~\ref{fig:BG_limit}, both models predict
similar trends and values. The estimated band gaps are mostly within
1eV below the experimental values. PPf is the only polymer showing a
PA-like behavior: the band gap converging to $0$ as $1/m$. This can
be explained by symmetry arguments \cite{Wennerstrom1985}. The
non-linear (in $1/m$) convergence of other oligomer series is in
line with experimental observations. The band gap decreases by
$1.5-2$eV between $m=2$ and $m=\infty$ in series of para-connected
oligomers PPP$m$ and PPV$m$, whereas for the meta-connected PMP$m$
and PmPV$m$ this decrease is only $\approx 0.5$eV, in a good
agreement with available experimental data. We have also compared
our results with Time Dependent Density Functional Theory (TD DFT)
calculation \cite{Ma2002}, data represented by triangles. These data
compare better with measurements than the FE or HMO predictions, but
provide less information about trends.

\begin{figure}
  \begin{center}
    \mbox{\hspace*{-1cm}
      \includegraphics[width=0.6\textwidth]{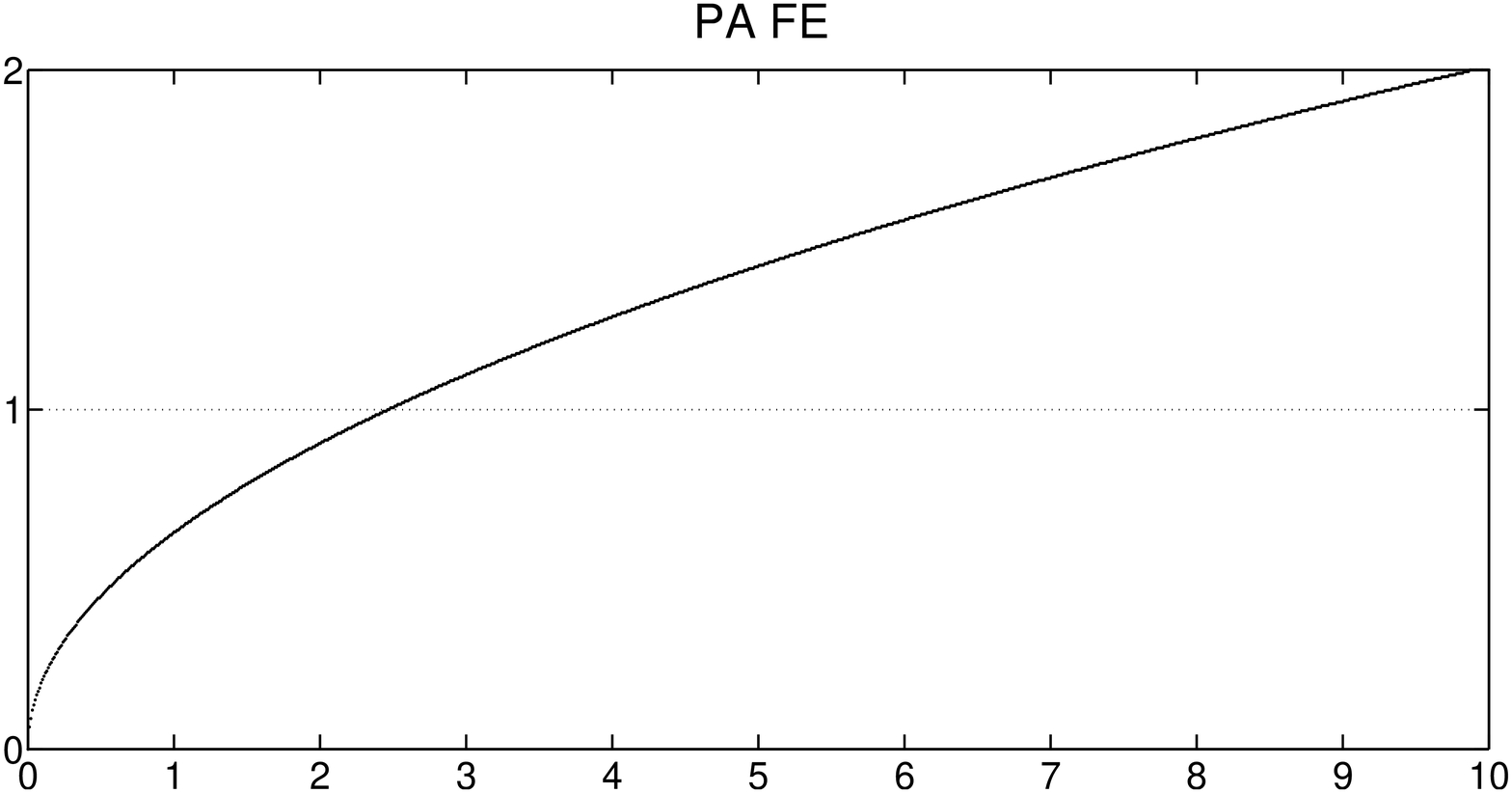}
      \includegraphics[width=0.6\textwidth]{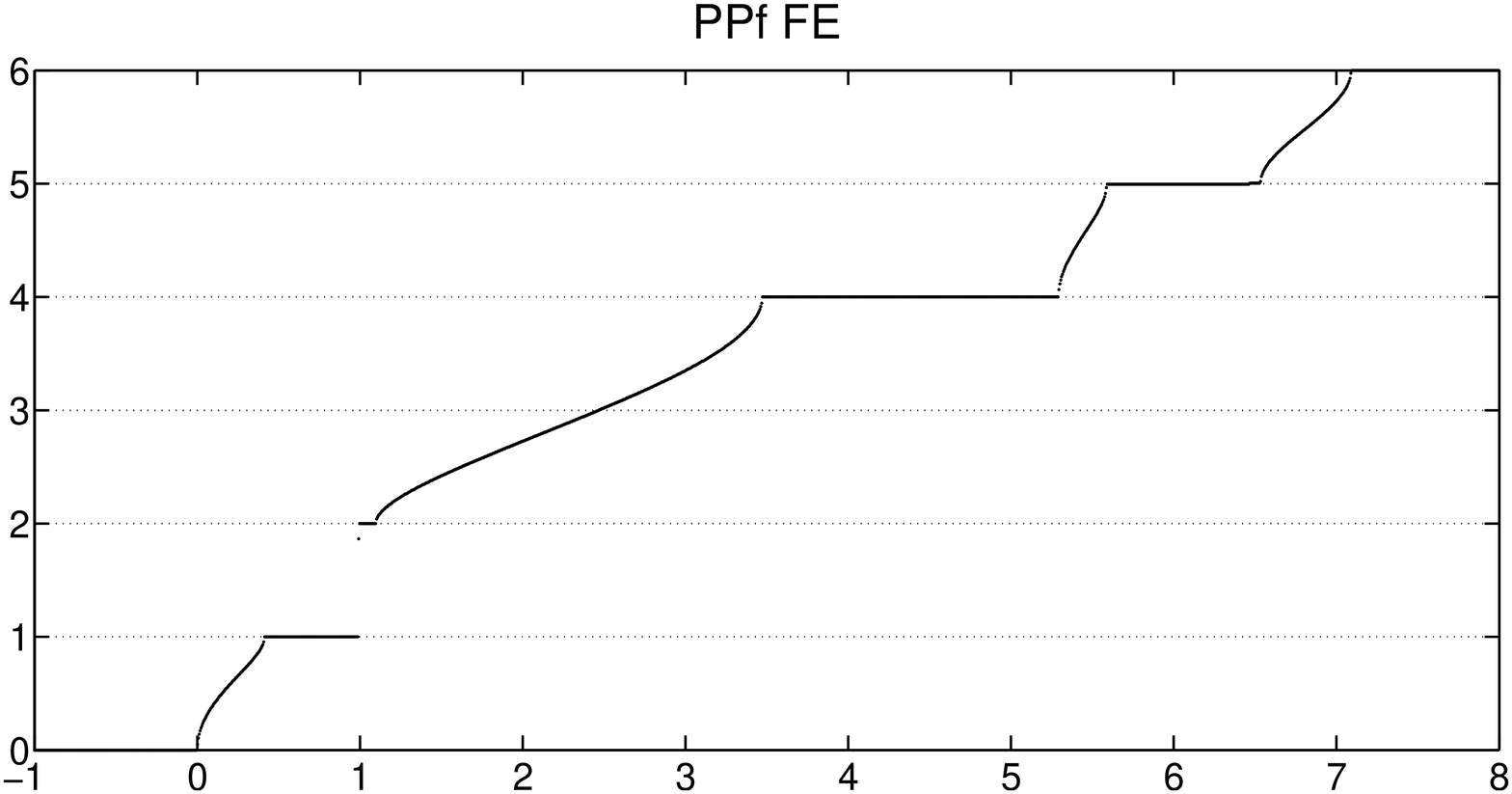}
      \hspace{-1cm}
    }
    \mbox{\hspace*{-1cm}
      \includegraphics[width=0.6\textwidth]{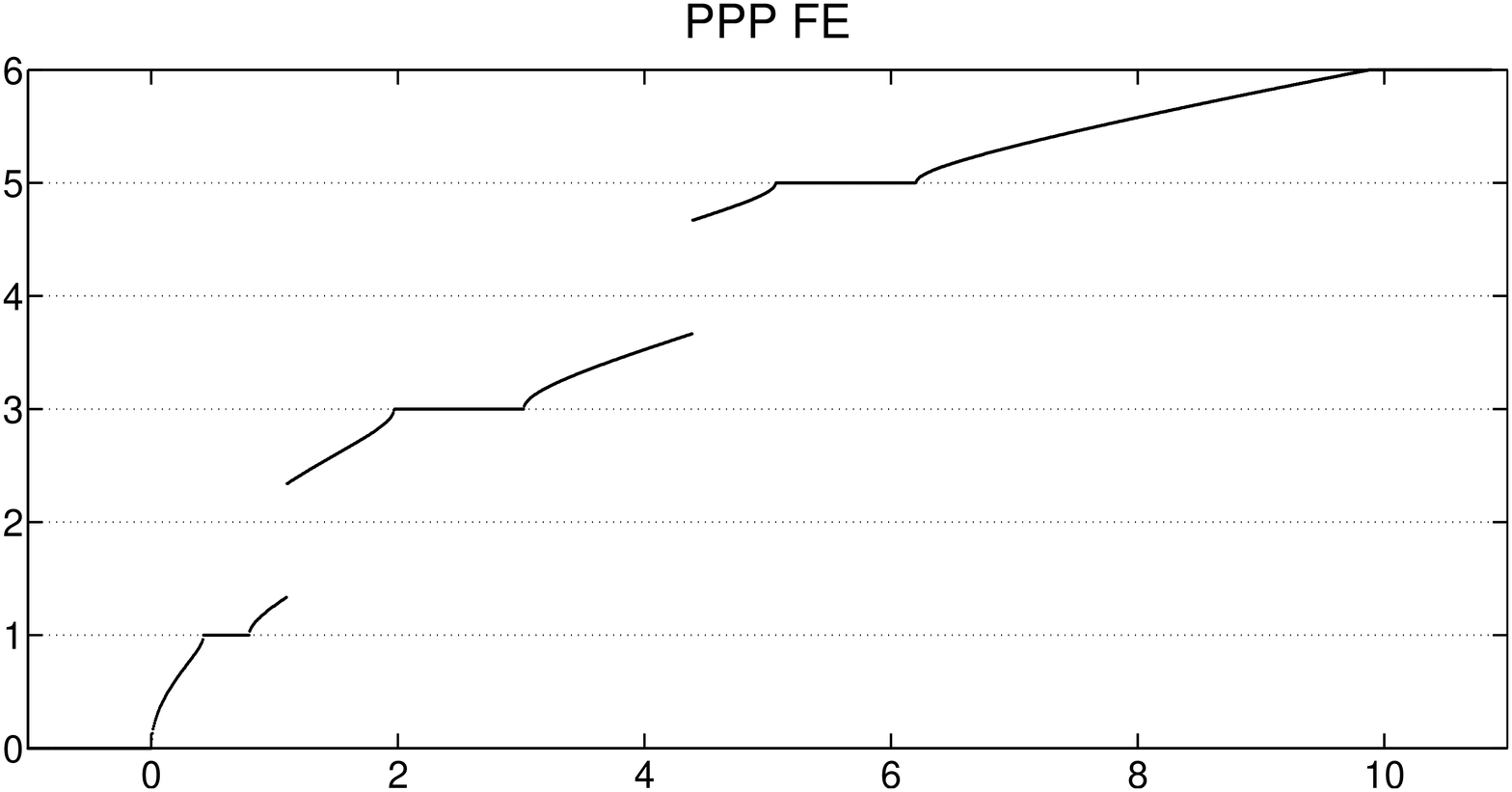}
      \includegraphics[width=0.6\textwidth]{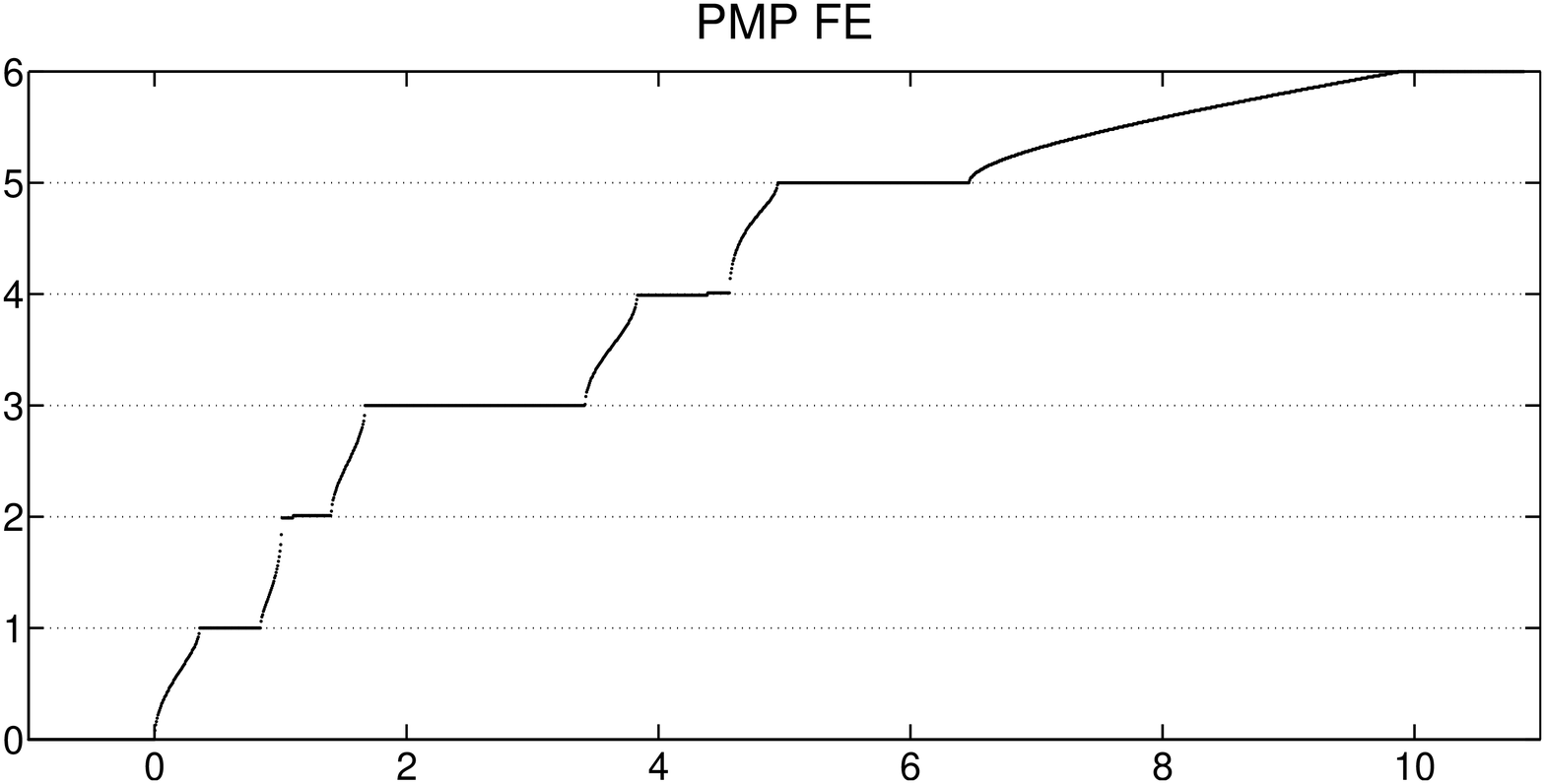}
      \hspace{-1cm}
    }
    \mbox{\hspace*{-1cm}
      \includegraphics[width=0.6\textwidth]{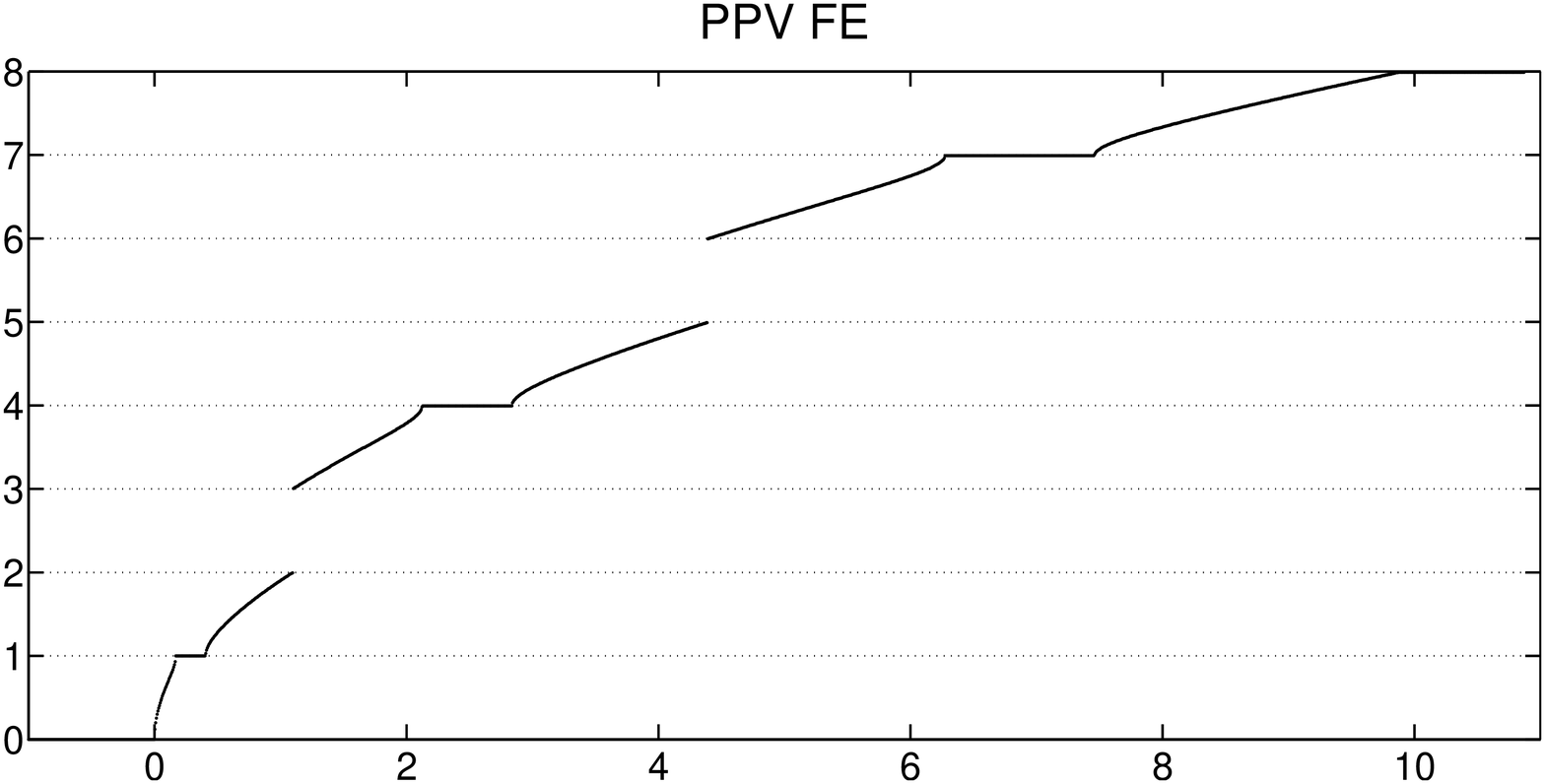}
      \includegraphics[width=0.6\textwidth]{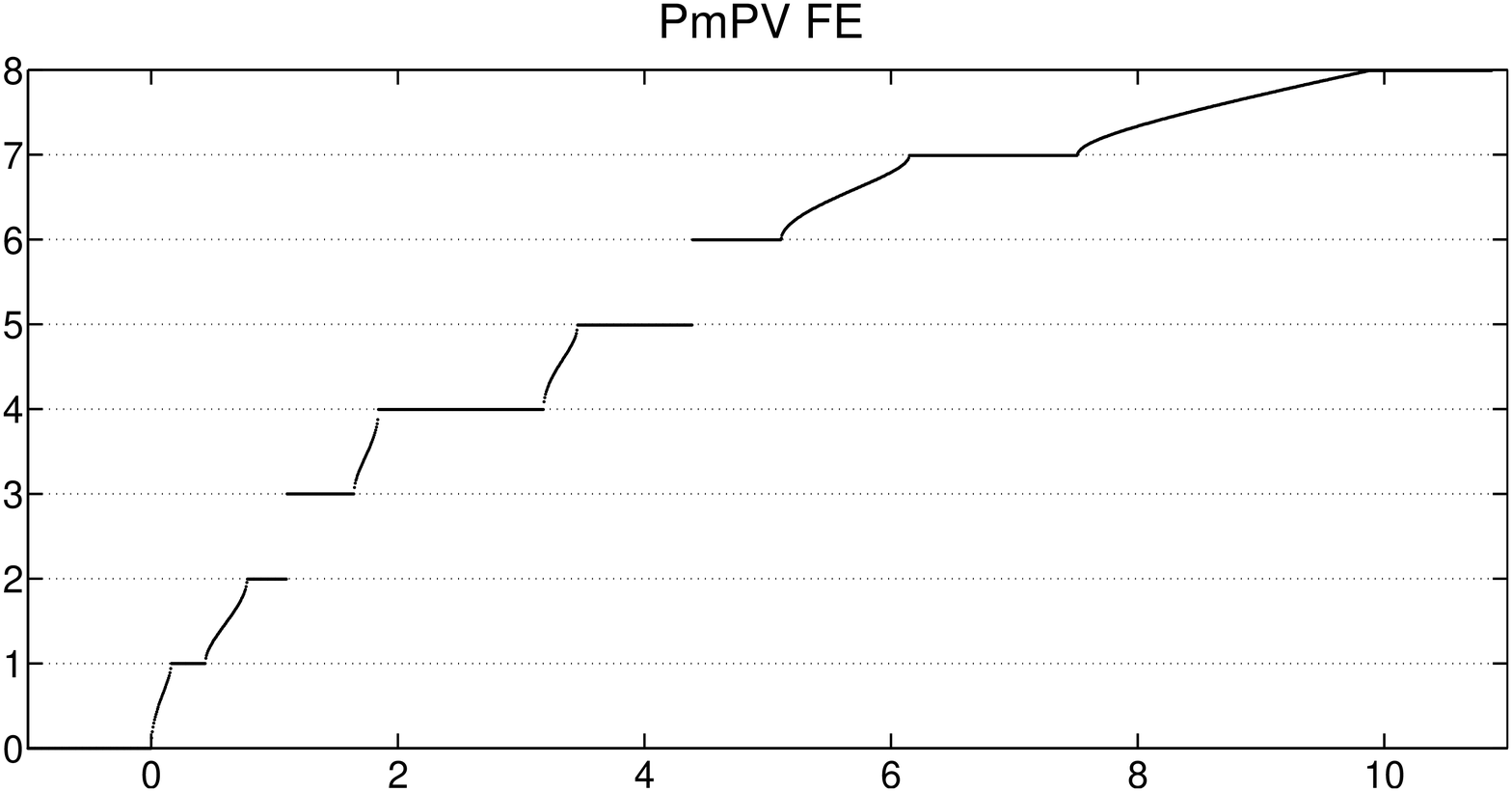}
      \hspace{-1cm}
    }
  \end{center}
  \caption{Spectral counting function $\sigma^{200}_{\rm FE}$
  of PA, PPf, PPP, PMP, PPV and PmPV in the FE model}\label{fig:bands_FE}
\end{figure}

\begin{figure}
  \begin{center}
    \mbox{\hspace*{-1cm}
      \includegraphics[width=0.6\textwidth]{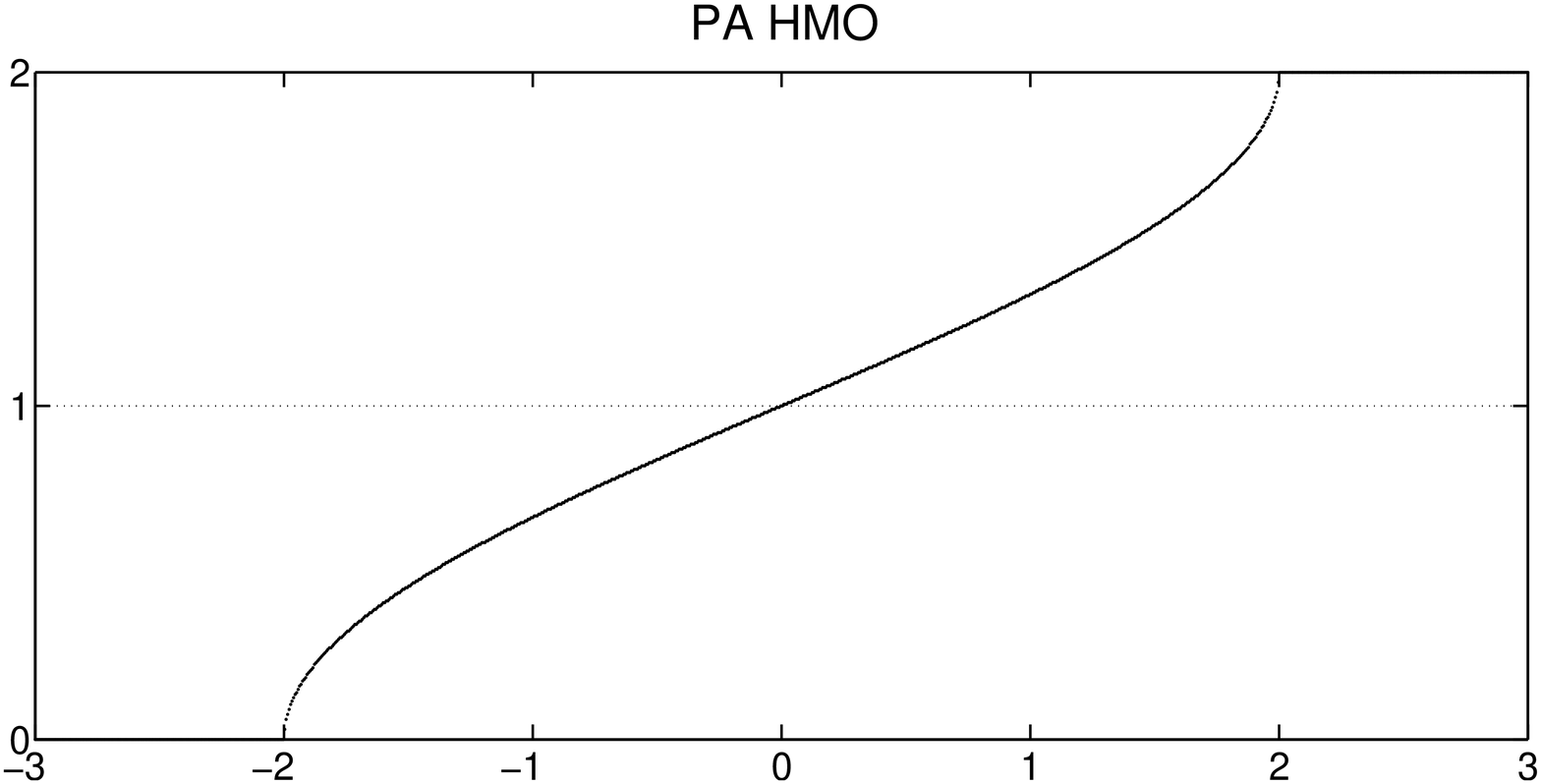}
      \includegraphics[width=0.6\textwidth]{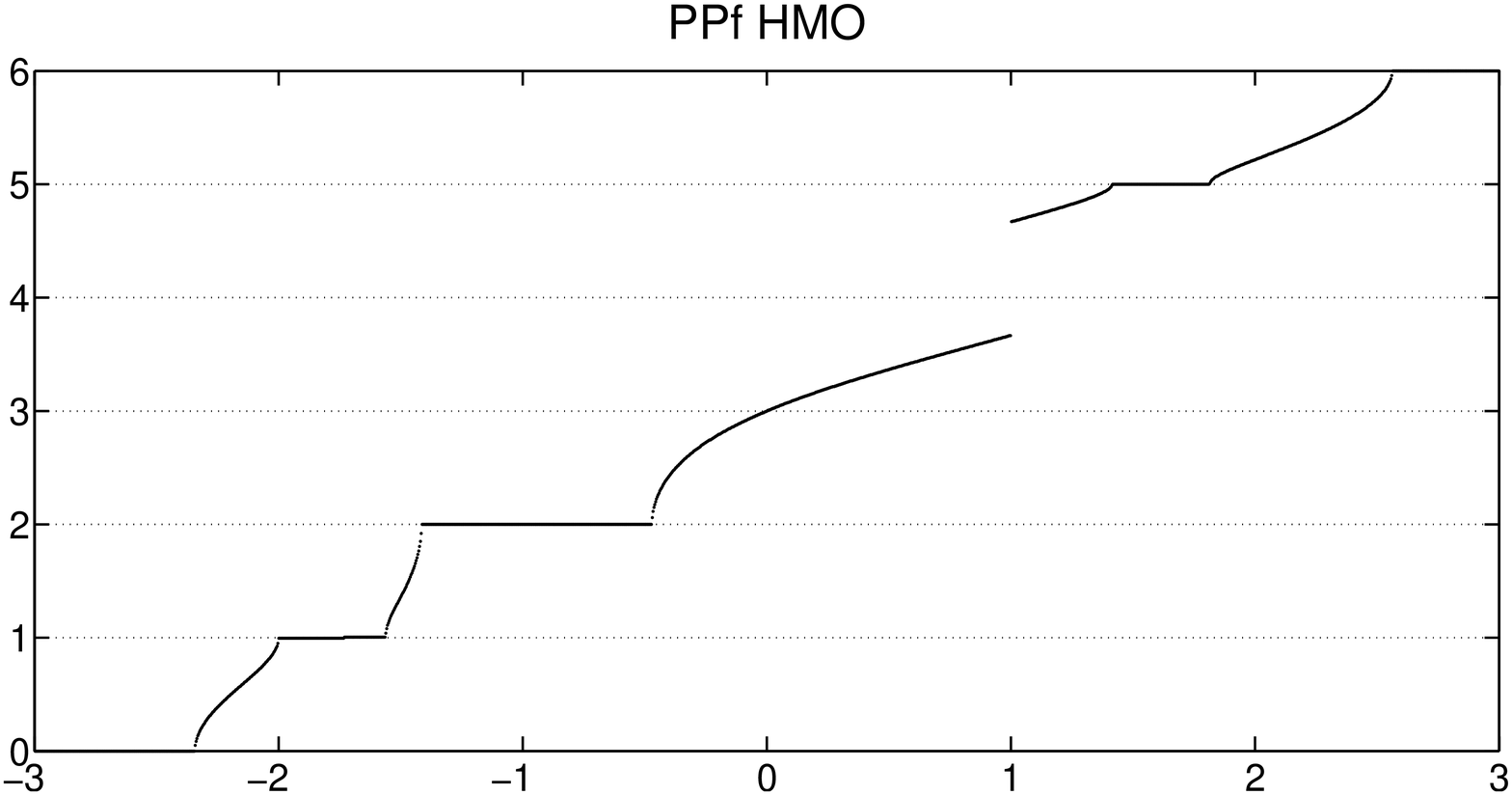}
      \hspace{-1cm}
    }
    \mbox{\hspace*{-1cm}
      \includegraphics[width=0.6\textwidth]{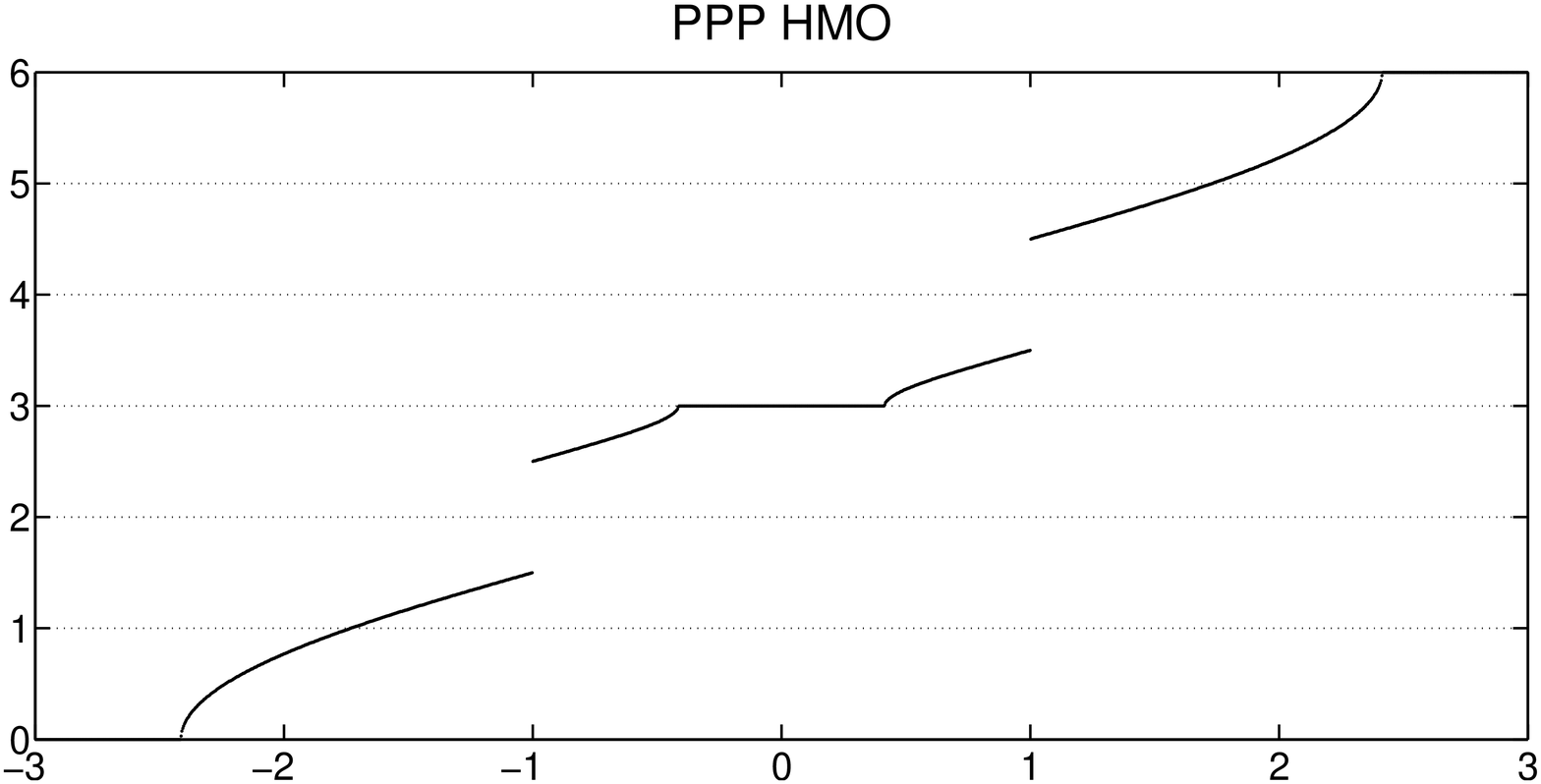}
      \includegraphics[width=0.6\textwidth]{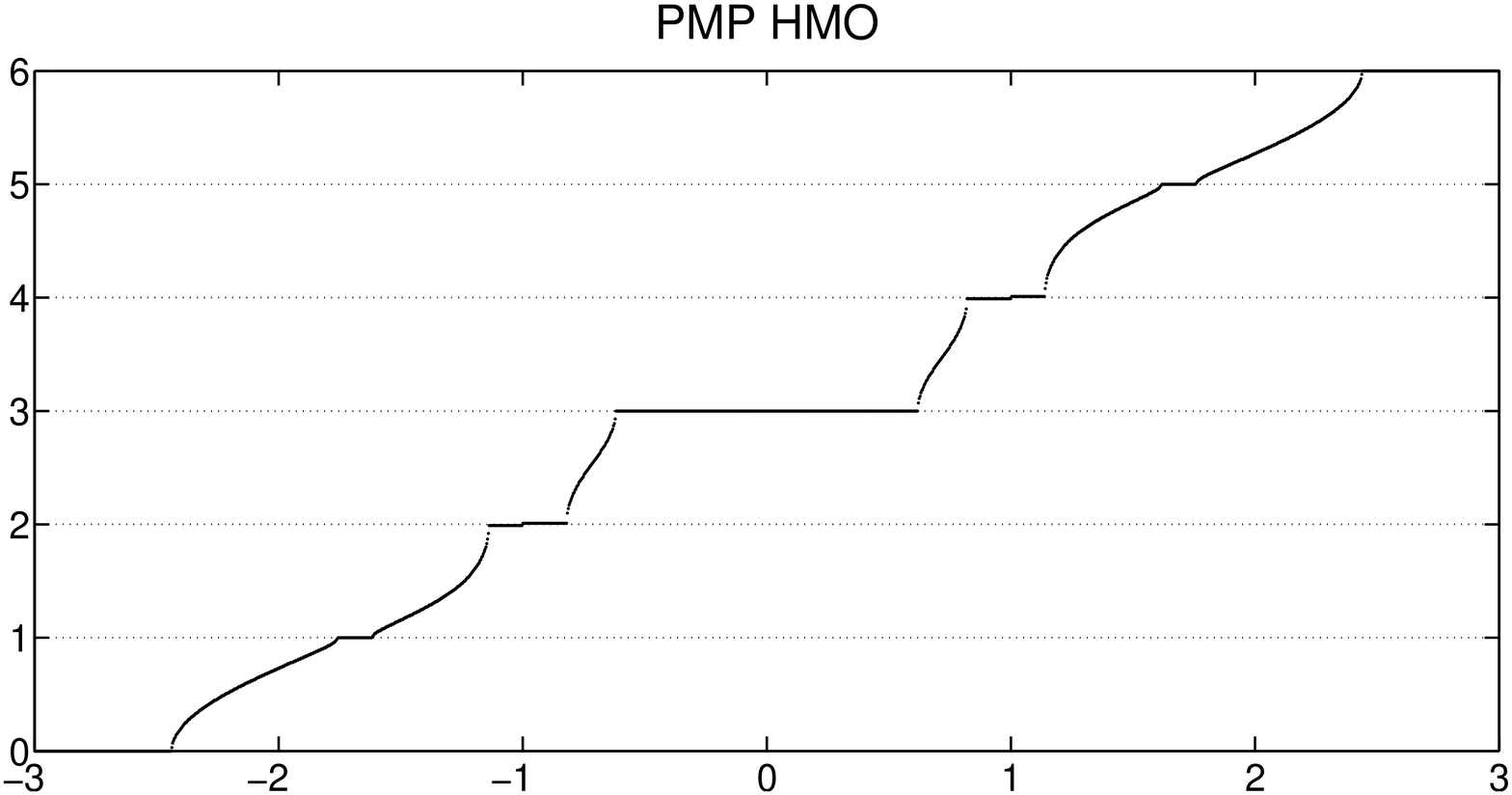}
    \hspace{-1cm}}
    \mbox{\hspace*{-1cm}
      \includegraphics[width=0.6\textwidth]{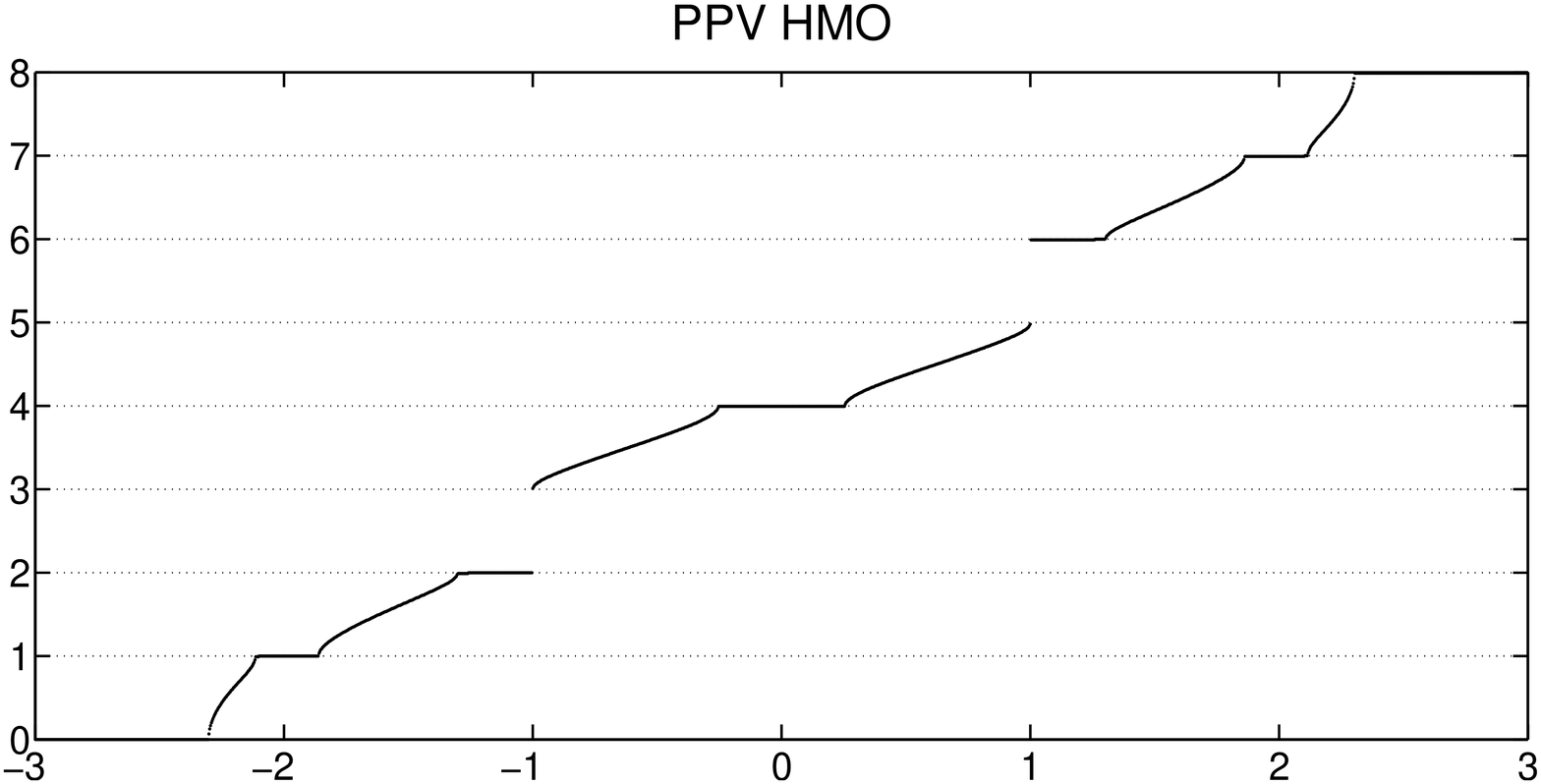}
      \includegraphics[width=0.6\textwidth]{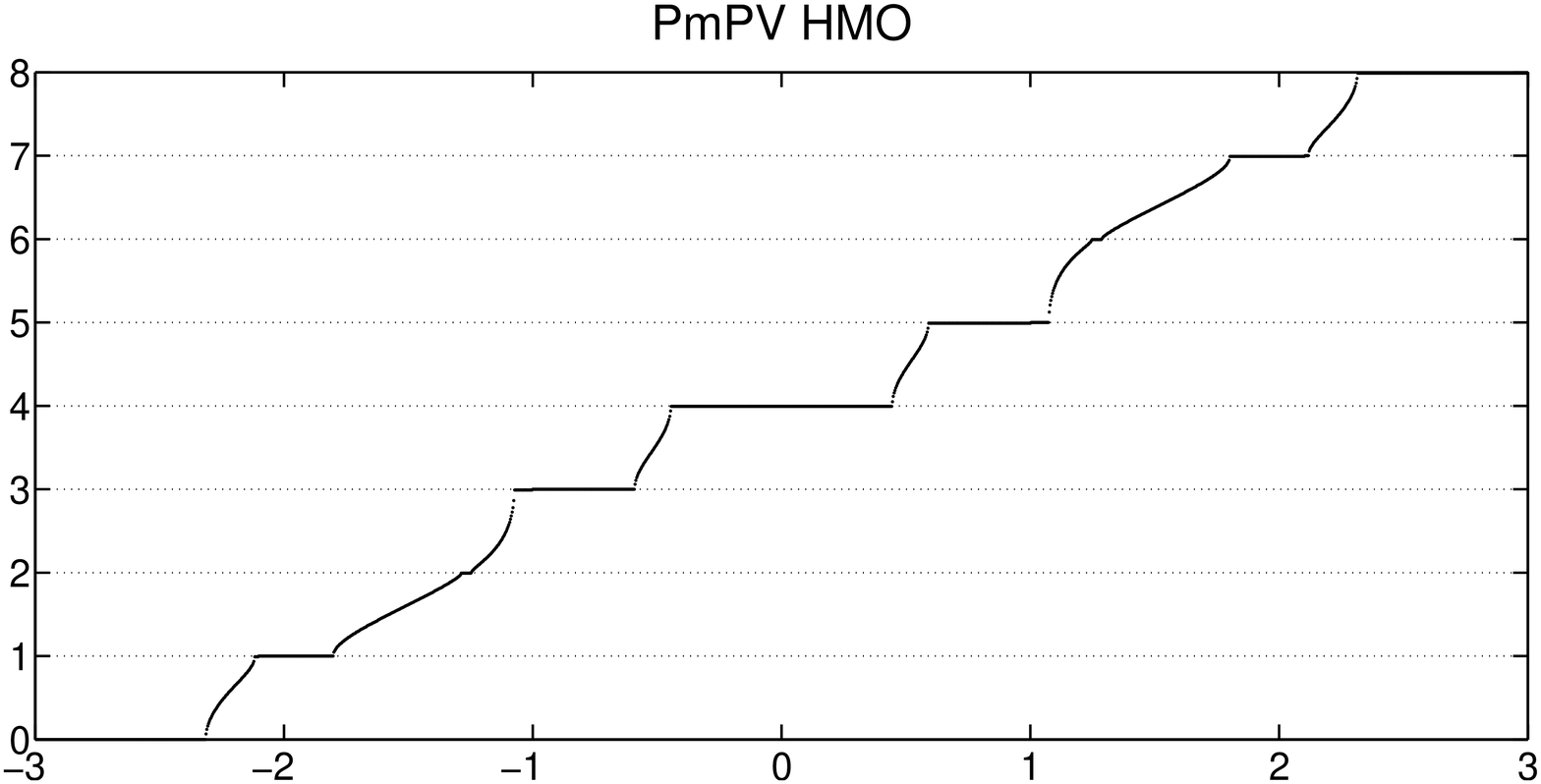}
    \hspace{-1cm}}
  \end{center}
  \caption{Spectral counting function $\sigma^{200}_{\rm HMO}$
  of PA, PPf, PPP, PMP, PPV and PmPV in the HMO model}\label{fig:bands_HMO}
\end{figure}

Another crucial property for conductivity is the width of the
valence and conduction bands. We have calculated $\sigma^m_{\rm
HMO}$ and $\sigma^m_{\rm FE}$ for the six above polymers ($m=200$)
in Figs.~\ref{fig:bands_FE} and \ref{fig:bands_HMO}.

As discussed above, the band gaps of PA and PPf vanish. We can see
that the para-connected polymers have much wider valence and
conduction bands than their meta-connected counterparts. Together
with the smaller band gaps this explains why PPP and PPV are good
conductors while PMP and PmPV are not \cite{Skotheim1986}.

\section{Conclusions}
We have studied conducting polymers from the point of view of H\"uckel Molecular
Orbital Theory and Free Electron Model, the two models
directly related to the spectral graph theory. Qualitative predictions match the
experimental reality. Floquet-Bloch theory for periodic
graphs provides means for an easy study of band properties of polymers.

Quantitative comparison shows that actual band gaps can be
reasonably estimated by either of the models when correcting for
intramolecular and intermolecular shifts. The absence of any
parameter fitting makes this result particularly exciting. The
convergence behavior $\triangle E(m) - \triangle E(\infty)$ of band
gaps of oligomer series is well reproduced. Therefore, it is
possible to extract the band gap of the polymer from the data for
finite values of $m$ with sufficient accuracy using simple graph
models. It would be promising for practical applications to derive
an analytic expression for $\triangle E(m)$.

\section*{Acknowledgements}
PS would like to thank the Condensed Matter and Interfaces group at
Utrecht University for hospitality.

\bibliography{lib_QG}
\bibliographystyle{plain}
\end{document}